\documentclass[sigconf,edbt]{acmart-edbt2018}

\usepackage{booktabs} 

\setcopyright{rightsretained}

\usepackage{graphicx}
\usepackage{balance}  
\usepackage{mathrsfs}
\usepackage[ruled, vlined, linesnumbered]{algorithm2e}
\usepackage{algorithmic}
\usepackage{amsmath,amssymb} 
\usepackage{subfig}
\usepackage{epsfig}
\usepackage{multirow}

\newtheorem{theo}{Theorem}

\acmDOI{}

\acmISBN{978-3-89318-078-3}

\acmConference[EDBT 2018]{21st International Conference on Extending Database Technology (EDBT)}{March 26-29, 2018}{Vienna, Austria} 
\acmYear{2018}

\begin{document}

\title{Sequenced Route Query  with Semantic Hierarchy}

\author{Yuya Sasaki$^{\dagger}$, Yoshiharu Ishikawa$^{\ddagger}$, Yasuhiro Fujiwara $^{\S\dagger}$, Makoto Onizuka$^{\dagger}$}
\affiliation{%
  \institution{$\dagger$Graduate School of Information Science and Technology, Osaka University, Osaka, Japan\\
  $\ddagger$Graduate School of Information Science, Nagoya University, Nagoya, Japan\\
  $\S$NTT Software Innovation Center, Tokyo, Japan
  }
}
\email{sasaki@ist.osaka-u.ac.jp, ishikawa@i.nagoya-u.ac.jp, fujiwara.yasuhiro@lab.ntt.co.jp, onizuka@ist.osaka-u.ac.jp}

\renewcommand{\shortauthors}{}

\begin{abstract}
The trip planning query searches for preferred routes starting from a given point through multiple Point-of-Interests (PoI) that match user requirements.   
Although previous studies have investigated trip planning queries, they lack flexibility for finding routes because all of them output routes that strictly match user requirements.
We study trip planning queries that output multiple routes in a flexible manner.
We propose a new type of query called {\it skyline sequenced route (SkySR)} query, which searches for all preferred sequenced routes to users by extending the shortest route search with the semantic similarity of PoIs in the route.
Flexibility is achieved by the {\it semantic hierarchy} of the PoI category.
We propose an efficient algorithm for the SkySR query, {\it bulk SkySR algorithm} that simultaneously searches for sequenced routes and prunes unnecessary routes effectively.
Experimental evaluations show that the proposed approach significantly outperforms the existing approaches in terms of response time (up to four orders of magnitude).
Moreover, we develop a prototype service that uses the SkySR query, and conduct a user test to evaluate its usefulness.
\end{abstract}

%
%


\maketitle

\section{Introduction}
Recently, technological advances in various devices, such as smart phones and automobile navigation systems, have allowed users to obtain real-time location information easily.
This has triggered the development of location-based services such as Foursquare, which exploit rich location information to improve service quality.
The users of the location-based services often want to find short routes that pass through multiple Points-of-Interest (PoIs); consequently, developing trip planning queries that can find the shortest routes that passes through user-specified categories has attracted considerable attention \cite{dai2016personalized,li2005trip}.  
If multiple PoI categories, e.g., restaurant and shopping mall, are in an ordered list (i.e., a {\it category sequence}), the trip planning query searches for a {\it sequenced route} that passes PoIs that match the user-specified categories in order. 

\begin{figure}[ttt]
	\centering
	\includegraphics[width=0.9\linewidth]{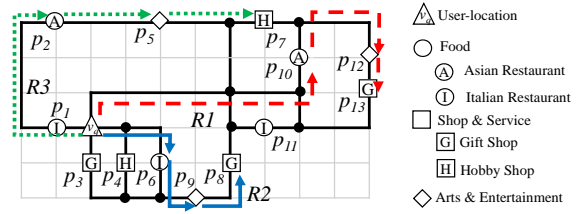}
	\caption{An example of a road network with PoIs}
	\label{fig:graph}
\end{figure}

\begin{example}\label{ex:intro}
Figure \ref{fig:graph} shows a road network with the following PoIs: ``Asian restaurant'', ``Italian restaurant'', ``Gift shop'', ``Hobby shop'', and ``Arts$\&$Entertainment (A$\&$E)''. 
Assume that a user wants to go to an Asian restaurant, an A$\&$E place, and a gift shop in this order from start point $v_q$.
The sequenced route query outputs route $R1$ because it is the shortest route from $v_q$ that satisfied the user requirements $\langle$Asian restaurant, A$\&$E, gift shop$\rangle$.
\end{example}

\begin{figure}[ttt]
	\centering
	\includegraphics[width=0.9\linewidth]{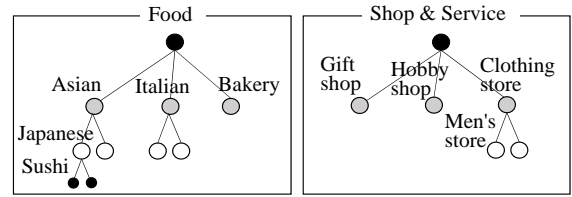}
	\caption{Examples of category trees in Foursquare}
	\label{fig:category}
\end{figure}

Existing approaches find the shortest route based on the user query.
However, such approaches may find an unexpectedly long route because the found PoIs may be distant from the start point.
A major problem with the existing approaches is that they only output routes that {\it perfectly} match the given categories \cite{eisner2012sequenced,ohsawa2012sequenced,sharifzadeh2008optimal}.
To overcome this problem, we introduce flexible similarity matching based on PoI category classification to find shorter routes in a flexible manner.
In the real-world, category classification often forms a {\it semantic hierarchy}, which we refer to as a {\it category tree}. 
For example, in Foursquare\footnote{https://developer.foursquare.com/categorytree}, the ``Food'' category tree includes ``Asian restaurant,'' ``Italian restaurant,'' and ``Bakery'' as subcategories, and the ``Shop $\&$Service'' category includes ``Gift shop,'' ``Hobby shop,'' and ``Clothing store'' as subcategories (Figure \ref{fig:category}).
We employ this semantic hierarchy to evaluate routes in terms of two aspects, i.e., route length and the semantic similarity between the categories of the PoIs in the route and those specified in the user query.
As a result, we can find effective sequenced routes that {\it semantically} match the user requirement based on the semantic hierarchy. 
For example, in Figure \ref{fig:graph}, route $R2$ satisfies the user requirement because it semantically matches the category sequence because Italian and Asian restaurants are in the same category tree.
However, this approach may find a significantly large number of sequenced routes because the number of PoIs that flexibly match the given categories increases significantly.
To reduce the number of routes to be output, we employ the skyline concept \cite{borzsony2001skyline}, i.e., we restrict ourselves to searching for the routes that are not worse than any other routes in terms of their scores (i.e., numerical values to evaluate the routes).
Based on this concept, we propose the {\it skyline sequenced route (SkySR) query}, which applies the skyline concept to the route length and semantic similarity (i.e., we consider route length and semantic similarity as route scores). Given a start point and a sequence of PoI categories, a SkySR query searches for sequenced routes that are no worse than any other routes in terms of length and semantic similarity.

\begin{example}
Table \ref{intro:result_ny} shows real-world examples of sequenced routes in New York city where a user plans to go to a cupcake shop, an art museum, and then a jazz club in this order.
The existing approaches output a single route that matches the user's requirement perfectly. 
The proposed approach can output three additional routes that are shorter than the route found by the existing approach. 
Note that the additional routes also satisfy the user query semantically. 
The user can select a preferred route among all the four routes depending on how far he/she does not want to walk or their available time.
\end{example}


\begin{table}[t]
	\begin{center}
	\caption{Example routes in New York city} 
		\label{intro:result_ny}
		\footnotesize	
		\begin{tabular}{|l|l|l|} \hline
Approach & Distance & Sequenced route	\\ \hline \hline
Existing &\multirow{2}{*}{3239 meters} & \multirow{2}{*}{Cupcake Shop $\rightarrow$ Art Museum $\rightarrow$ Jazz Club}\\ 
 (e.g., \cite{sharifzadeh2008optimal}) & &\\ \hline
\multirow{4}{*}{Proposed}&3239 meters & Cupcake Shop $\rightarrow$ Art Museum $\rightarrow$ Jazz Club\\ 
&1858 meters & Dessert Shop $\rightarrow$ Art Museum $\rightarrow$ Jazz Club\\
&1392 meters & Dessert Shop $\rightarrow$ Museum $\rightarrow$ Jazz Club\\
&823 meters & Dessert Shop $\rightarrow$ Museum $\rightarrow$ Music Venue \\
\hline
		\end{tabular}
        \end{center}
\end{table}
\begin{table*}[t]
	\begin{center}
	\caption{Types of trip planning queries.} 
	\label{table:query}
	\footnotesize	
		\begin{tabular}{|l|r|r|r|r|r|r|} \hline
		Type							 &  Distance metrics 					&	Order		& Destination  	& Result	& Scores\\\hline	
		SkySR (proposed)  & Network & Total & Yes or No & Exact  & Length and semantic \\
		Optimal sequenced route (OSR)  \cite{sharifzadeh2008optimal}				 & Euclidean or Network & Total & Yes or No & Exact & Length \\
		Sequenced route \cite{eisner2012sequenced, ohsawa2012sequenced}				 & Network & Total & Yes & Exact & Length \\
		Personalized sequenced route \cite{dai2016personalized}	& Euclidean & Total & No & Approximate & Length and rating \\ 
		Trip planning \cite{li2005trip}	 & Euclidean or Network & Non & Yes & Approximate & Length \\
		Multi rule partial sequenced route \cite{chen2008multi}			 & Euclidean & Partial & No & Approximate & Length \\
		Multi rule partial sequenced route \cite{li2013optimal}			& Euclidean & Partial & No & Exact & Length\\
		Multi-type nearest neighbor  \cite{ma2006exploiting} & Euclidean & Non & No & Exact & Length\\\hline	
\end{tabular}
\end{center}
\end{table*}

The SkySR query can provide effective trip plans; however, it incurs significant computational cost because a large number of routes can match the user requirement.
Therefore, the SkySR query requires an efficient algorithm. 
The challenge is to search for SkySRs efficiently by reducing the search space without sacrificing the exactness of the result.
We propose {\it bulk SkySR} algorithm ({\sf BSSR} for short) that finds exact SkySRs efficiently.
Recall that a feature of SkySRs is that their scores are no worse than those of other sequenced routes. 
{\sf BSSR} exploits the branch-and-bound algorithm \cite{lawler1966branch}, which effectively prunes unnecessary routes based on the upper and lower bounds of route scores. 
In addition, to improve efficiency more, we employ four techniques to optimize {\sf BSSR}. (1) First, we initially find sequenced routes to calculate the upper bound. (2) We tighten the upper bound by arranging the priority queue and (3) tighten the lower bound by introducing minimum distances. 
(4) we keep intermediate results for later processing, which refer to as {\it on-the-fly caching}.
Our approach significantly outperforms existing approaches in terms of response time (up to four orders of magnitude) without increasing memory usage or sacrificing the exactness of the result.

The main contributions of this paper are as follows.
\begin{itemize}
\item We introduce a semantic hierarchy to the route search query, which allows us to search for routes flexibly. 
\item We propose the {\it skyline sequenced route (SkySR) query}, which finds all preferred routes related to a specified category sequence with a semantic hierarchy (Section \ref{sec:def}). 
\item We propose an exact and efficient algorithm and its optimization techniques to process SkySR queries (Section \ref{sec:pro}). 
\item We discuss variations and extensions of the SkySR query. The SkySR query can be applied to various user requirements and environments (Section \ref{sec:dis}).
\item We demonstrate that the proposed approach works well in terms of response time and memory usage by performing extensive experiments. (Section \ref{sec:exp}). 
\item We develop a prototype service that employs the SkySR query and conduct a user test to evaluate usefulness of the SkySR query. (Section \ref{sec:usertest}).
\end{itemize}

The remainder of this paper is organized as follows. Section \ref{sec:rel} introduces related work.  Section \ref{sec:pre} describes the problem formulation, and Section \ref{sec:def} defines the SkySR query. Section \ref{sec:pro} presents the proposed algorithm. In Section \ref{sec:dis}, we discuss variations and extensions of the SkySR query. Sections \ref{sec:exp} and \ref{sec:usertest} present experiment and user test results, respectively, and Section \ref{sec:con} concludes the paper.

\section{Related work}
\label{sec:rel}

First, we review trip planning query studies related to the SkySR query.
Then, we review some studies related to the skyline operator.
To the best of our knowledge, no study has considered a skyline sequenced route; thus, our problem cannot be solved efficiently using existing approaches.

{\bf Trip planning:} We categorize trip planning queries in Table \ref{table:query}.
Note that all existing trip planning queries only output routes that perfectly match the user-specified category sequences.
Moreover, since most trip planning queries assume Euclidean distance, they cannot  find SkySRs, in which road network distance is assumed. 
Dai et al.\ \cite{dai2016personalized} proposed a personalized sequenced route and assumed that PoIs have ratings as well as categories and that users assign weighting factors as preferences. 
Although this personalized sequenced route considers route lengths and ratings, it only outputs the route that perfectly matches the given categories and has the best score based on lengths, ratings, and preferences. 
Only the optimal sequenced route (OSR) is applicable to find SkySRs without modification because the OSR and SkySR are based on the same settings (except for scoring).
Sharifzadeh et al.\ \cite{sharifzadeh2008optimal} proposed two algorithms to find OSRs in road networks: the {\it Dijkstra-based solution} and the {\it Progressive Neighbor Exploration (PNE) approach}.
The main difference between these algorithms is that the Dijkstra-based solution employs the Dijkstra algorithm to search for PoIs and the PNE approach employs the nearest neighbor search.
It has been reported that these algorithms are comparable in terms of performance \cite{sharifzadeh2008optimal}.
Thus, we consider both algorithms to verify the performance of the proposed approach.

{\bf Skyline:} The skyline operator was proposed previously \cite{borzsony2001skyline}. 
Few studies have considered the skyline concept for route searches.
Recently, the skyline route (or skyline path) has received considerable attention \cite{aljubayrin2015skyline,hansen1980bicriterion, kriegel2010route, martins1984multicriteria,shekelyan2015linear,tian2009finding, yang2014stochastic}.
A skyline route assumes that edges on road networks are associated with multiple costs, such as distance, travel time, and tolls.
Here, the objective is to find skyline routes from a start point to a destination considering these multiple costs.
However, since we specify a category sequence rather than a destination, we cannot apply conventional algorithms to find SkySRs.
The continuous skyline query in road networks (e.g., \cite{huang2005route}) searches for the skyline PoIs for a moving object considering both the PoI category and the distances to the moving object.
Because continuous skyline queries search for a single PoI category, these solutions are not applicable to SkySR queries, which obtain routes that pass through multiple PoIs. 

\begin{table}[t]
	\begin{center}
	\caption{Notations}
	\label{symbols}
			\footnotesize	
		\begin{tabular}{|l|l|} \hline
Symbol&Meaning	\\ \hline 
$\mathbb{V}$ & Set of vertices \\
$\mathbb{P}$ & Set of PoI vertices \\
$\mathbb{E}$ & Set of edges \\
$p$ & PoI vertex \\
$\mathbb{C}$ & Set of categories \\
$c$ & Category \\
$t$  & Category tree \\
$c_p$ & Category of PoI vertex $p$ \\
$t_c$ & Category tree of $c$ \\
$\mathbb{P}_c$ & Set of PoI vertices associated with $c$\\
$\mathbb{P}_t$ & Set of PoI vertices associated with $t$\\
$\mathbf{S}$& Category sequence (sequence of categories) \\
$\mathbf{R}$& Route (sequence of PoI vertices)\\
$\mathbf{S}_R$& Sequential PoI categories in $\mathbf{R}$\\
$l(\mathbf{R})$& Length score of $\mathbf{R}$\\
$s(\mathbf{R})$& Semantic score of $\mathbf{R}$\\
$\mathcal{R}$& Set of routes \\
$\mathcal{E}(\mathbf{R})$& Set of super-routes of $\mathbf{R}$ \\
$\mathcal{S}$& Minimal set of sequenced routes \\
$\mathbf{S}_q$& Category sequence specified by user \\
$v_q$ & Start point specified by user  \\\hline
		\end{tabular}
        \end{center}
\end{table}

\section{Preliminaries}
\label{sec:pre}
Table \ref{symbols} summarizes the notations used in this paper.
We assume a connected graph $G=(\mathbb{V} \cup \mathbb{P}, \mathbb{E})$, where $\mathbb{V}$, $\mathbb{P}$, and $\mathbb{E}  \subseteq (\mathbb{V}\cup \mathbb{P}) \times (\mathbb{V}\cup \mathbb{P})$ represent the sets of vertices, PoI vertices, and edges, respectively.
This graph corresponds to a road network that contains PoIs.
The numbers of vertices, PoI vertices, and edges are denoted $|\mathbb{V}|$, $|\mathbb{P}|$, and $|\mathbb{E}|$, respectively.
PoI vertex $p \in \mathbb{P}$ is associated with category $c \in \mathbb{C}$, where $\mathbb{C}$ is the set of categories.
We denote the category of PoI vertex $p$ as $c_p$, and assume that each PoI is associated with a single category.
Each category is associated with category tree $t$, and we denote the category tree of category $c$ as $t_c$.
We denote the set of PoI vertices associated with $c$ and the set of PoI vertices associated with the category tree $t$ as $\mathbb{P}_c$ and $\mathbb{P}_t$, respectively.
If a PoI vertex is associated with category $c$, it is also associated with all ancestor categories of $c$ in $t_c$.
Each edge $e(u_i, u_j)$ in $\mathbb{E}$ is associated with a weight $w(u_i,u_j)~(\geq 0)$. The weight can represent either travel duration or distance.
Next, we define several terms required to introduce the skyline sequenced route (SkySR).

\begin{definition}{\bf (Category sequence)}
A {\it category sequence} $\mathbf{S} = \langle c_S[1], c_S[2], \ldots , c_S[|\mathbf{S}|] \rangle$ is a sequence of categories, where $|\mathbf{S}|$ is the size of $\mathbf{S}$.
$c_S[i] \in \mathbb{C}$ denotes the $i$-th category in $\mathbf{S}$. A {\it super-category sequence} of $\mathbf{S}$ is a category sequence where each $i$-th category is either $c_S[i]$ or an ancestor of $c_S[i]$ ($1\leq i \leq |\mathbf{S}|$) in the category tree.
\end{definition}

\begin{definition}{\bf (Route)}
A {\it route} $\mathbf{R} = \langle p_R[1], \ldots , p_R[|\mathbf{R}|] \rangle$ is a sequence of PoI vertices in a road network, where $p_R[i] \in \mathbb{P}$ and $|\mathbf{R}|$ denote the $i$-th PoI vertex in $\mathbf{R}$ and the size of $\mathbf{R}$, respectively.
$\mathbf{S}_R$ denotes the category sequence of $\mathbf{R}$ (i.e., $\langle c_{p_R[1]}, \ldots , c_{p_R[|\mathbf{R}|]} \rangle$). 
In addition, we define a {\it super-route} of $\mathbf{R}$ as an extended route of $\mathbf{R}$, such as $\langle \mathbf{R}, p_i, p_j, \ldots \rangle$.
In other words, a super-route of $\mathbf{R}$ is obtained by adding a sequence of PoI vertices to the end of $\mathbf{R}$.
$\mathcal{R}$ and $\mathcal{E}(\mathbf{R})$ denote a set of routes and a set of super-routes of $\mathbf{R}$, respectively.
Moreover, given a route  $\mathbf{R} = \langle p_R[1], \ldots , p_R[|\mathbf{R}|] \rangle$ and a PoI vertex $p$, we define $\mathbf{R} \oplus p = \langle p_R[1], \ldots , p_R[|\mathbf{R}|], p \rangle$. 
\end{definition}

\begin{definition}{\bf (Category similarity)}
Given two categories $c$ and $c'$, the similarity $sim(c, c')~\in [0, 1]$ is calculated by an arbitrary function such as the \emph{Wu and Palmer similarity} or path length \cite{resnik1995using, wu1994verbs}.
We assume the following relations in the similarity. 
\begin{itemize}
\item $c$ is irrelevant to $c'$ if both exist in different category trees; thus, we obtain  $sim(c, c')=0$. 
\item $c$ {\it semantically matches} $c'$ if $c$ and $c'$ are in the same category tree; thus, we obtain  $0 < sim(c, c') \leq 1$. 
\item $c$ {\it perfectly matches} $c'$ if $c$ and $c'$ are the same; thus, we obtain $sim(c, c')=1$. 
\end{itemize}
Note that a semantic match subsumes a perfect match.
\end{definition}

We define a {\it sequenced route} using the above definitions.
The difference between our definition of {\it sequenced route} and the previous definition \cite{sharifzadeh2008optimal} is that we consider category similarity.

\begin{definition}{\bf (Sequenced route)}\label{def:seqencedroute}
Given category sequence  $\mathbf{S}= \langle c_S[1], \ldots , c_S[|\mathbf{S}|] \rangle $, $\mathbf{R} = \langle p_R[1], \ldots , p_R[|\mathbf{R}|] \rangle $ is {\it a sequenced route} of category sequence $\mathbf{S}$ if and only if it satisfies (i) $|\mathbf{R}|=|\mathbf{S}|$, (ii) $c_S[i]$ semantically matches $c_{p_R[i]}$ for all $i$ such that $1\leq i\leq |\mathbf{S}|$, and (iii) all PoI vertices in $\mathbf{R}$ differ each other.
\end{definition}

\begin{definition}{\bf (Route scores)}\label{def:score}
Given category sequence  $\mathbf{S}$ and vertex $v$ as a start point, we define two scores for route $\mathbf{R}$:  {\it length score} $l(\mathbf{R}) \in [0, \inf]$ and {\it semantic score} $s(\mathbf{R}) \in [0, 1]$. We define the length score $l(\mathbf{R})$ as follows:
\begin{equation}\label{eq:length}
l(\mathbf{R})=  D(v, p_R[1]) + \Sigma_{i=1}^{|\mathbf{R}|-1} D(p_R[i], p_R[i+1]),
\end{equation}
where $D(u_i, u_j)$ denotes the smallest weight sum of the edges on the routes between vertices (or PoIs) $u_i$ and $u_j$. The semantic score $s(\mathbf{R})$ is calculated by an aggregation function $f$ as follows:
\begin{equation}\label{eq:semantic}
s(\mathbf{R})= f(h_1, h_2, \ldots, h_{|\mathbf{R}|}),
\end{equation}
where $h_i$ denotes $sim(c_S[i], c_{p_R[i]})$. We assume that, if all $h_i =1$, $s(\mathbf{R}) = 0$, i.e., if all PoI vertices in a route perfectly match the categories, the semantic score of the given route is 0. 
We also assume that $\underbar{s}(\mathbf{R})$ is the possible minimum semantic score of $\mathbf{R}$ when it is a sequenced route.
Without loss of generality, preferred routes have small length and semantic score.
\end{definition}

\section{The skyline sequenced route query}
\label{sec:def}

Here, we define the SkySR query.
Intuitively, a SkySR is a potential route that may be the best route related to the user's requirement.
A potential route is a route that is not {\it dominated} by any other routes; the notion of \emph{dominance} is used in the {\it skyline operator} \cite{borzsony2001skyline}.  
We define dominance for sequenced routes and SkySR query in the following. 
\begin{definition}{\bf (Dominance)}\label{def:sky}
Let $\mathcal{R}$ be the set of all sequenced routes starting from point $v$ for category sequence $\mathbf{S}$.
For two sequenced routes $\mathbf{R}, \mathbf{R}' \in \mathcal{R}$, we say that $\mathbf{R}$ dominates $\mathbf{R}'$ if we have (i) $l(\mathbf{R})\! < \!l(\mathbf{R}')$ and  $s(\mathbf{R})\! \leq \! s(\mathbf{R}')$ or (ii) $s(\mathbf{R})\! <\! s(\mathbf{R}')$ and $l(\mathbf{R})\! \leq \!l(\mathbf{R}')$.
If two sequenced routes have the same length and semantic scores, the routes are {\it equivalent} in the dominance, and a set of sequenced routes is {\it minimal} if it has no equivalent routes.
\end{definition}

\begin{definition}{\bf (SkySR query)}\label{def:skysr}
Given vertex $v_q$ as a start point  and category sequence $\mathbf{S}_q$, a skyline sequenced route is a sequenced route not dominated by other routes.
Let  $\mathcal{R}$  be the set of all sequenced routes from start point $v_q$ for category sequence $\mathbf{S}_q$, and let $\mathcal{S}$ be a {\it minimal} set of the sequenced routes.
The SkySR query returns $\mathcal{S}$ that includes sequenced routes such that all $\mathbf{R} \in \mathcal{S}$ are SkySRs and all $\mathbf{R}' \in \mathcal{R} \setminus \mathcal{S}$ are dominated by or equivalent to some of $\mathbf{R} \in \mathcal{S}$.
\end{definition}

An naive solution to find SkySRs is to first enumerate SkySR candidates by iteratively executing OSR queries for any super-category sequences of $\mathbf{S}_q$ and then check the dominance among the routes.
The number of super-category sequences of $\mathbf{S}_q$ increases exponentially as the depth of the category in the category tree and the size of $\mathbf{S}_q$ increase.
Thus, although OSR algorithms can find a sequenced route efficiently, we must repeat many searches. 
As a result, the naive solution needs significantly high computational cost to find SkySRs.

\section{Proposed Algorithm}
\label{sec:pro}
In this section, we present the proposed approach, which we refer to as the {\it bulk SkySR algorithm}  ({\sf BSSR}), that finds SkySRs efficiently.
Section \ref{sec:design} presents the {\sf BSSR} design policy, and Section \ref{sec:BSSR} explains the {\sf BSSR} procedure. In Section \ref{sec:opt}, we propose optimization techniques for {\sf BSSR}.
We also theoretically analyze its performance in Section \ref{sec:cost}.
Finally, we show a running example of {\sf BSSR} in Section \ref{sec:runex}.
In Section \ref{sec:pro}, we assume undirected graphs in which each PoI vertex is associated with only one category and that users give sequences of single PoI categories.
However, in a real application, the graphs would be directed graphs, each PoI vertex would be associated with multiple categories, and users may specify complex categories.
Section \ref{sec:dis} describes how we handle the above conditions.

\subsection{Design Policy}
\label{sec:design}
Our idea to improve efficiency is to find sequenced routes simultaneously  (i.e., by searching sequenced routes in bulk) in order to reduce the search space.
We have two choice as the basis for our approach;  Dijkstra-based or nearest neighbor-based approaches \cite{sharifzadeh2008optimal}.
We use the Dijkstra-based approach as the basis of our algorithm.
Recall that a SkySR query has two scores for a route, i.e., length and semantic scores. 
To find all SkySRs, we must find routes that have small category scores even if the routes have large length scores.
However, PoIs that are included in the routes with small category scores could be distant from the start point. 
Although the nearest neighbor-based approach finds the closest PoIs, it cannot efficiently find such PoIs.
On the other hand, the Dijkstra-based approach searches for all PoI vertices that match a PoI category.
Therefore, the Dijkstra-based approach is more suitable for the SkySR query than the nearest neighbor-based approach.

Although our approach  finds sequenced routes simultaneously, it entails a large number of executions of the Dijkstra algorithm.
This is  because, since the number of PoI candidates increases, a large number of possible routes increases.
The search space does not become small effectively.
To effectively reduce the search space, we exploit the branch-and-bound algorithm, which uses the upper and lower bounds of a branch of the search space to solve an optimization problem effectively.
With {\sf BSSR}, each branch corresponds to each route.
For the upper and lower bounds, we compute the bounds during finding the set of SkySRs.
Specifically, we compute the upper bound of a route from the already found sequenced routes, and we compute the lower bound from the current searched route (i.e., not a sequenced route yet).
With the upper and lower bounds, we can safely prune unnecessary routes to improve efficiency. 

To further increase efficiency, we propose optimization techniques for {\sf BSSR}.
In order to  exploit the branch-and-bound algorithm, it is necessary to initialize the upper bound.
Thus, we first search for a sequenced route to initialize the upper bound. However, it may take high computational cost to find a sequenced route.
Therefore, we propose a {\it nearest neighbor-based initial search method} ({\sf NNinit}) that finds sequenced routes efficiently by greedily finding PoI vertices.
In addition, to effectively update the upper bound, we assign a priority to each route and use the priority queue to efficiently find routes that are likely to give an effective upper bound.
To compute the lower bound, we compute the possible minimum distance and add it to the length score of a route to safely prune unnecessary routes.
Moreover,  to avoid executing the Dijkstra algorithm iteratively from the same vertices, we materialize search results of the Dijkstra algorithm and reuse them to search the PoI vertices.
By using {\sf BSSR} with optimization techniques, we can perform the SkySR query efficiently.

\subsection{Bulk SkySR algorithm}
\label{sec:BSSR}

{\it Bulk SkySR algorithm} ({\sf BSSR}) finds all SkySRs by finding simultaneously sequenced routes with checking dominance on demand.
The naive solution must execute OSR queries for all super-category sequences of $\mathbf{S}_q$ one by one because it only searches for the PoIs that perfectly match the given category. 
In contrast, {\sf BSSR} searches for all PoIs that semantically match the given category.  

The basic process of {\sf BSSR} is simple as shown in Algorithm \ref{alg:bulk_ssr}: (i) start searching the PoI vertices that match the first category from start point $v_q$ and insert the route  found into priority queue $Q_b$ which stores all found routes (line 4), (ii) fetch a route from $Q_b$ (line 6), (iii) search for the next PoI vertices that semantically match the next category $c_d$ from PoI vertex $p_d$ which is the end of the fetched route, and insert the fetched route with each of the found PoI vertices into $Q_b$ (lines 7--9), and (iv) if $Q_b$ is not empty, return to (ii), otherwise output the minimal set of sequenced route $\mathcal{S}$ (line 10). In steps (i) and (iii), we find  PoI vertices from the end of the fetched route using a Dijkstra algorithm modified for the SkySR query as described in Section \ref{sssec:modDijk}.

\begin{algorithm}[h]
	{\footnotesize
	\caption{Bulk SkySR algorithm}	\label{alg:bulk_ssr}
		\DontPrintSemicolon

			    \SetKwFunction{BSSR}{BSSR}
			    \SetKwFunction{PrunedDijkstra}{mDijkstra}
			      \SetKwFunction{NNbasedEst}{NNmethod}
	            \SetKwInOut{Input}{input}\SetKwInOut{Output}{output}
            	{\bf procedure} \BSSR{$v_q$, $\mathbf{S}_q$}\\
	                   $\mathcal{S} \leftarrow$ $\phi$; \\
		                priority$\_$queue $Q_{b}$ $\leftarrow$ $\phi$;\\
		                \PrunedDijkstra{$\phi$, $c_S[1]$, $v_q$, $Q_{b}$,  $\mathcal{S}$};\\     
                        \While{$Q_{b}$ is not empty }{
		                       $\mathbf{R}$ $\leftarrow$ $Q_{b}$.dequeue();\\
		                       $c_d$ $\leftarrow$ $c_S[|\mathbf{R}|+1]$;\\
		                       $p_d$ $\leftarrow$ $p_R[|\mathbf{R}|]$;\\
		                      \PrunedDijkstra{$\mathbf{R}$, $c_d$, $p_d$, $Q_{b}$,  $\mathcal{S}$};\\                    
                        }
                  return $\mathcal{S}$;\\
                  {\bf end procedure}
                  }
\end{algorithm}

\subsubsection{Branch-and-bound}
\label{sec:branch}
We search for sequenced routes simultaneously to reduce the search space.
Our idea to safely reduce the search space is to exploit the branch-and-bound algorithm, which can reduce unnecessary search space.
This section describes the theoretical background of using the branch-and-bound algorithm. 
We use the following three lemmas to reduce the search space:

\begin{lemma}{}\label{lemma:upper}
Let $\mathcal{S}$ be a minimum set of sequenced routes while searching for SkySRs and $\mathcal{S}'$ be the minimum set of sequenced routes after finding SkySRs.
If sequenced route $\mathbf{R}$ is dominated by a sequenced route in $\mathcal{S}$, $\mathbf{R}$ cannot be included in  $\mathcal{S}'$.
\end{lemma}
{\it proof:} From Definition \ref{def:skysr}, we search for a set of SkySRs, which are not dominated by the other sequenced routes.
If we find a sequenced route not dominated by any sequenced routes in $\mathcal{S}$, we update $\mathcal{S}$ by inserting the new sequenced route and deleting a sequenced route dominated by the new one.
Therefore, any sequenced routes in $\mathcal{S}$ after the update are not dominated by any sequenced routes in $\mathcal{S}$ prior to the update.
As a result, sequenced routes in $\mathcal{S}'$ are not dominated by any sequenced routes in $\mathcal{S}$.
In other words, $\mathbf{R}$ is not included in $\mathcal{S}'$ if we have sequenced route $\mathbf{R}'$ in $\mathcal{S}$ such that $l(\mathbf{R}') \leq l(\mathbf{R})$ and $s(\mathbf{R}') \leq s(\mathbf{R})$. \hfill{} $\square$

\begin{lemma}{}\label{lemma:lower}
Let $\mathcal{E}(\mathbf{R})$ be a set of super-routes of $\mathbf{R}$ starting from the same start point.
For any route $\mathbf{R}'$ in $\mathcal{E}(\mathbf{R})$, the length and semantic scores $l(\mathbf{R}')$ and $s(\mathbf{R}')$ cannot be less than $l(\mathbf{R})$ and $\underbar{s}(\mathbf{R})$, respectively.
\end{lemma}
{\it proof:} Let $\mathbf{R}'$ be a route included in $\mathcal{E}(\mathbf{R})$. 
Since we have  $D(u_i, u_j) \geq 0$, the following property holds for a route $\mathbf{R}$ from Equation (\ref{eq:length}) of Definition \ref{def:score}.
\begin{eqnarray*}
 & \!\!\!\!\!\!& \!\!\!\!\!\! D(v_q, p_{R'}[1])  +  \Sigma_{i=1}^{|\mathbf{R}'|-1} D(p_{R'}[i], p_{R'}[i\!+\!1])  \; \\ 
 & \!\!\!\!\!\! =  & \!\!\!\!\!\!       D(v_q, p_{R}[1])   +     \Sigma_{i = 1}^{|\mathbf{R} |  -1} \! D(p_{R}[i], p_{R}[i\!  +\!  1]) \! \\
 &&\hspace{3cm} +   \Sigma_{i  = |\mathbf{R}|}^{|\mathbf{R'}|-1} \!D(p_{R'}[i], p_{R'}[i\!  +\!  1]) \; \\
 & \!\!\!\!\!\! \geq  & \!\!\!\!\!\! D(v_q, p_{R}[1]) + \Sigma_{i=1}^{|\mathbf{R}|-1} D(p_{R}[i], p_{R}[i\!+\!1]).\; 
\end{eqnarray*}

\noindent
Therefore, we have $l(\mathbf{R}) \leq l(\mathbf{R}')$.
$\underbar{s}(\mathbf{R})$ is the possible minimum semantic score of $\mathbf{R}$ when it becomes a sequenced route.
Thus, even if PoI vertices are added to $\mathbf{R}$, we have  $\underbar{s}(\mathbf{R}) \leq s(\mathbf{R}')$.
As a result, we have   $l(\mathbf{R}) \leq l(\mathbf{R}')$ and  $\underbar{s}(\mathbf{R}) \leq s(\mathbf{R}')$. \hfill{}$\square$

In terms of the branch-and-bound algorithm, Lemma \ref{lemma:upper} and \ref{lemma:lower} give us the upper and lower bounds of the scores of a route, respectively.
We can prune routes according to the following lemma.

\begin{lemma}{}\label{lemma:prune}({\bf pruning condition})
If (i) $\mathbf{R}$ is a sequenced route included in the set $\mathcal{S}$ of sequenced routes and (ii) $l(\mathbf{R}) \leq l(\mathbf{R}')$ and  $s(\mathbf{R}) \leq \underbar{s}(\mathbf{R}')$, any routes in $\mathcal{E}(\mathbf{R}')$ cannot be included in $\mathcal{S}$.
\end{lemma}
{\it proof:} If we have $l(\mathbf{R}) \leq l(\mathbf{R}')$ and  $s(\mathbf{R}) \leq \underbar{s}(\mathbf{R}')$, $\mathbf{R}'$ is not included in $\mathcal{S}$ (Lemma \ref{lemma:upper}).
From Lemma \ref{lemma:lower}, the scores of $\mathbf{R}'$ cannot become less than $l(\mathbf{R}')$ and $\underline{s}(\mathbf{R}')$ even if we expand $\mathbf{R}'$.
Therefore, any routes in $\mathcal{E}(\mathbf{R}')$ cannot be included in $\mathcal{S}$ because $\mathbf{R}'$ is dominated by or equivalent to the sequenced route with $l(\mathbf{R})$ and $s(\mathbf{R})$ . \hfill{}$\square$

Lemma \ref{lemma:prune} gives us the length score {\it threshold} for a route, and, if the length score of a route is greater than this threshold, we can prune the given route.
We define the length score threshold of a route as follows:

\begin{definition}
The threshold $\overline{l}(\mathbf{R})$ of the length score of route $\mathbf{R}$ is given by the following equation:
\begin{equation}\label{eq:up}
\overline{l}(\mathbf{R})= \min_{\mathbf{R}' \in \mathcal{S}} \{  l(\mathbf{R}') | \underbar{s}(\mathbf{R}) \geq s(\mathbf{R}') \}.
\end{equation} 
\end{definition}
If $\overline{l}(\mathbf{R}) \leq l(\mathbf{R})$, we can safely prune $\mathbf{R}$ because it cannot be included in the result. 
Thus, we can reduce the search space without sacrificing the exactness of the result. 
Equation (\ref{eq:up}) has a small computation cost because $\mathcal{S}$ includes only a small number of sequenced routes as shown in Section \ref{sec:exp}.

\subsubsection{The modified Dijkstra Algorithm}
\label{sssec:modDijk}
We search the next PoI vertices that semantically match the next PoI category using the modified Dijkstra algorithm.
The modified Dijkstra algorithm can prune unnecessary routes based on Lemma \ref{lemma:prune}.
Moreover, based on the following lemma, it terminates unnecessary traversal of the graph and avoids inserting unnecessary routes.

\begin{lemma}{}\label{lemma:property}
Let $\mathbf{R} = \langle p_R[1], \ldots , p_R[i], p_R[i+1], p_R[i+2] \ldots ,$ $p_R[|\mathbf{R}|] \rangle$ be a route and $p_{i:i+1}$ be a PoI vertex on a path between $p_R[i]$ and $p_R[i+1]$. 
Route $\mathbf{R}$ must be dominated by or equivalent to another route if we have $sim(c_S[i+1], c_{p_{i:i+1}}) \geq sim(c_S[i+1], c_{p_R[i+1]})$.
\end{lemma}
{\it proof:} Let $\mathbf{R}' = \langle p_R[1], \ldots , p_R[i], p_{i:i+1}, p_R[i+2], \ldots, $ $p_R[|\mathbf{R}|] \rangle$ be a route such that the difference between $\mathbf{R}$ and $\mathbf{R}'$ is only in $p_{i:i+1}$ and  $p_R[i+1]$.
Since the PoI vertex $p_{i:i+1}$ is on the path between $p_R[i]$ and $p_R[i+1]$, we have $l(\mathbf{R}) \geq l(\mathbf{R}')$ based on triangle inequality (i.e., $D( p_{i:i+1}, p_{R}[i + 1])+D(p_{R}[i+1], p_{R}[i + 2]) \geq D( p_{i:i+1}, p_{R}[i + 2])$ ).
Moreover, if $sim(c_S[i+1], c_{p_{i:i+1}}) \geq sim(c_S[i+1], c_{p_R[i+1]})$, we have $s(\mathbf{R}) \geq s(\mathbf{R}')$.
Therefore,  $\mathbf{R}$ is dominated by or equivalent to  $\mathbf{R}' $ because $l(\mathbf{R}) \geq l(\mathbf{R}')$ and  $s(\mathbf{R}) \geq s(\mathbf{R}')$. \hfill{}$\square$

Lemma \ref{lemma:property} gives us two properties for the SkySR query: (i) even if we find a PoI vertex that passes through another PoI vertex that has a better category similarity, we can ignore the PoI vertex, and (ii) if we find a PoI vertex that perfectly matches the given category, we do not need to traverse the graph through the PoI vertex.
As a result, using Lemma \ref{lemma:prune} and \ref{lemma:property}, we can efficiently find the next PoI vertices.

Algorithm \ref{alg:next_poi} shows the pseudocode for the modified Dijkstra algorithm, which is used to find PoI vertices that semantically match $c_d$ from $p_d$.
In priority queue $Q_{d}$ for the modified Dijkstra algorithm, the top vertex is the closest vertex to $p_d$.
The queue is initialized to $p_d$ (line 3).
The closest vertex to $p_d$ is dequeued from $Q_{d}$ (line 5).
$\mathbf{R}_t$ is a route expanded from $\mathbf{R}_d$, which is $\mathbf{R}_d$ with fetched vertex $u$ (line 7).
If the length score of $\mathbf{R}_t$ is greater than or equal to the threshold of $\mathbf{R}_d$, the modified Dijkstra algorithm terminates the process (Lemma \ref{lemma:prune}) (line 8).
We check whether (i) $u$ semantically matches $c_d$ and (ii) $u$ does not proceed through another PoI vertex whose category similarity is greater than or equal to that of $u$ (line 9).
If we satisfy the above conditions and the length score of $\mathbf{R}_t$ is less than its threshold (line 10), we insert $\mathbf{R}_t$ into the priority queue or the set of sequenced routes (lines 10--12).
Otherwise, we skip the process to insert  $\mathbf{R}_t$ (Lemma  \ref{lemma:prune} and \ref{lemma:property}).
The neighbor vertices of $u$ are inserted into $Q_{d}$ unless $u$ perfectly matches $c_d$ (Lemma \ref{lemma:property}) (lines 13--17).
\begin{algorithm}[h]
	{\footnotesize
	\caption{Modified Dijkstra algorithm to find the next PoI vertices matching $c_d$ from $p_d$}	\label{alg:next_poi}
		\DontPrintSemicolon
			    \SetKwFunction{MinUB}{MinUB}
			   \SetKwFunction{Dijkstra}{Dijkstra}
			      \SetKwFunction{PrunedDijkstra}{mDijkstra}
			       \SetKwFunction{upperbound}{ub}
	            \SetKwInOut{Input}{input}\SetKwInOut{Output}{output}
	            {\bf procedure} \PrunedDijkstra{$\mathbf{R}_d$, $c_d$, $p_d$, $Q_{b}$,  $\mathcal{S}$}\\  
	              $dist[u] = \inf$ for all $u \in \mathbb{V} \cup \mathbb{P}$, $dist[p_d]=0$;\\   
	              priority$\_$queue $Q_{d}$ $\leftarrow$ $\{p_d \}$;\\
 		          \While{$Q_{d}$ is not empty}{
 					   $u$ $\leftarrow$ $Q_{d}$.dequeue;\\
 					    \lIf{$u$ is already visited}{continue;}	
 					    $\mathbf{R}_t \leftarrow$ $ \mathbf{R}_d \oplus u$;\\
 					   \lIf{$l(\mathbf{R}_t) \geq \overline{l}(\mathbf{R}_d)$}{break;}
 					  
 					        \If{$u \in \mathbb{P}_{t_{c_d}}$ {\bf and} $u$ is not through the PoI vertex whose category similarity is higher than that of $u$}{
 					        \If{$l(\mathbf{R}_t) <  \overline{l}(\mathbf{R}_t)$}{ 
 					        	   \lIf{$\mathbf{R}_t$ is a sequenced route}{
 					        	  $\mathcal{S}$.update($\mathbf{R}_t$);
 					        	             }				                  
 					        	                  \lElse{ $Q_{b}$.enqueue($\mathbf{R}_t$);}
 						      }  
 					        }
 					         \If{$u \notin  \mathbb{P}_{c_d}$}{
 					          \For{each $u'$ for $e(u,u') \in \mathbb{E}$}{
   							         \If{$dist[u] +  w(u,u') < dist[u]$ }{
   								          $dist[u']=dist[u] +  w(u,u').w$;\\
   								          $Q_{d}$.enqueue($u'$); \\
   								          }
   						         }
 					         } 					         				            
 					        }
 			{\bf end procedure}
		                      
                        }
\end{algorithm}

\subsection{Optimization techniques}
\label{sec:opt}
In this section, we propose four optimization techniques for {\sf BSSR}.
Section \ref{sec:init} explains an initial search for sequenced routes and proposes {\sf NNinit}.
We then explain tightening the upper and the lower bounds in Section \ref{sec:priority_queue} and Section \ref{sec:lower}, respectively.
Furthermore, in Section \ref{sec:cache} we propose an {\it on-the-fly caching technique} to reuse previous search results of the modified Dijkstra algorithm.

\subsubsection{Initial search}
\label{sec:init}

We prune unnecessary routes efficiently using the branch-and-bound algorithm.
However, we cannot calculate the threshold of $\mathbf{R}$ if there are no sequenced routes in $\mathcal{S}$ whose semantic scores are not greater than that of $\underbar{s}(\mathbf{R})$ based on Equation  (\ref{eq:up}).
Therefore, initially, we search for the sequenced route whose semantic score is 0.
However, the length score of the sequenced route can be large if its semantic score is 0.
To tighten the threshold, we also search for sequenced routes whose semantic scores are greater than 0 because the length scores of them are less than that of the sequenced route with a semantic score of 0. We initially find several sequenced routes to tighten the upper bound.

We propose {\sf NNinit}, which searches for several sequenced routes efficiently.
{\sf NNinit} performs a nearest neighbor search repeatedly to find PoI vertices that perfectly match the given categories.
With this process, we can find a sequenced route whose semantic score is 0.
Moreover, {\sf NNinit} can find the PoI vertex that semantically matches the given category during the nearest neighbor search.
When we find the last visited PoI vertex, we may find PoI vertices that semantically match the last category in $\mathbf{S}_q$.
Therefore, we can obtain sequenced routes whose semantic scores are greater than 0 and length scores are small.
As a result, {\sf NNinit} can find several sequenced routes without incurring additional cost, and one of the sequenced routes has a semantic score of 0.

We present the pseudocode for {\sf NNinit} in Algorithm \ref{alg:NN-based}.
Here, priority queue $Q$ is initialized to start point $v_q$ (line 3).
{\sf NNinit} repeats the Dijkstra algorithm $|\mathbf{S}_q|$ times to find sequenced routes (line 4).
The Dijkstra algorithm is executed to search for the closest PoI vertex that perfectly matches $c_{S_q}[i]$ from the initial vertex  (the first initial vertex is $v_q$) (lines 5--19).
Here, the closest vertex to the initial vertex is dequeued from $Q$ (line 7).
If the algorithm finds a PoI vertex that perfectly matches  $c_{S_q}[i]$, this vertex is added to $\mathbf{R}$ and $Q$ is initialized to the PoI vertex (lines 12--15). 
When it finds the last PoI vertex that semantically matches $c_{S_q}[|\mathbf{S}_q|]$, it inserts the sequenced route into $\mathcal{S}$ (lines 9--11).
Finally, we obtain a set of sequenced routes, and one of the sequenced routes in $\mathcal{S}$ has a semantic score of 0. 

\begin{algorithm}[h]
{\footnotesize
	\caption{Initial search for finding sequenced routes with a small cost}	\label{alg:NN-based}
	            \SetKwFunction{NNbasedEst}{NNinit}
	            \SetKwInOut{Input}{input}\SetKwInOut{Output}{output}
	            {\bf procedure} \NNbasedEst{$v_q$, $\mathbf{S}_q$}\\
			          $\mathcal{S}$ $\leftarrow$ $\phi$,
			          $\mathbf{R}$ $\leftarrow$ $\phi$;\\       
		              priority$\_$queue $Q$ $\leftarrow$ $\{ v_q \}$;\\
		              \tcc{execute Dijkstra algorithm $|\mathbf{S}_q|$ times }
		              \For{ $i$ : 1 to $|\mathbf{S}_q|$ }{
				              $dist[u] = \inf$ for all $u \in \mathbb{V} \cup \mathbb{P}$, $dist[Q.top]=0;$ \\       
					          \While{$Q$ is not empty}{
						          $u$ $\leftarrow$ $Q$.dequeue;\\
						          \lIf{$u$ is already visited}{continue}				          					         
						          \If {$i=|\mathbf{S}_q|$ {\bf and}  $u \in \mathbb{P}_{t_{c_{S_q}[i]}}$}{
							              $\mathbf{R}' \leftarrow $ $\mathbf{R} \oplus u$;  \\
						     	          $\mathcal{S}$.update($\mathbf{R}'$);\\ 						     
						          } 
						          \If {$u \in \mathbb{P}_{c_{S_q}[i]}$}{
						         $\mathbf{R} \leftarrow$ $ \mathbf{R} \oplus u$; \\
						         $Q$ $\leftarrow$ $\{ u \}$; \\
						          break;
						          }
						          \For{each $u'$ for $e(u,u') \in E$}{
							         \If{$dist[u] +  w(u,u') < dist[u']$ }{
								          $dist[u']=dist[u] +  w(u,u')$;\\
								          $Q$.enqueue($u'$); \\
								          }
						         }
				            }

	               }
	               {\bf return} $\mathcal{S}$;\\
	               {\bf end procedure}
}
\end{algorithm}

\begin{example}
We show an example of {\sf NNinit} using Example \ref{ex:intro}, which searches an Asian restaurant, an A$\&$E place, and a gift shop in this order from start point $v_q$.
{\sf NNinit} executes the Dijkstra algorithm three times because the size of category sequence is three.
First, {\sf NNinit} searches PoI vertices that perfectly match Asian restaurant from $v_q$.
Then, it finds $p_2$ that is the closest PoI that perfectly match Asian restaurant to $v_q$.
Next, it searches the closest PoI vertex that perfectly matches A$\&$E to $p_2$ and then finds $p_5$.
From the next search, {\sf NNinit} inserts sequenced routes to $\mathcal{S}$ when it finds PoI vertices that semantically match gift shop.
{\sf NNinit} finds $p_7$ whose category is Shop$\&$Service (i.e., semantically match) and thus inserts $\langle p_2, p_5, p_7\rangle$ to  $\mathcal{S}$.
After finding $p_7$, it finds $p_8$ that perfectly matches gift shop and inserts $\langle p_2, p_5, p_8\rangle$ to  $\mathcal{S}$.
Finally {\sf NNinit} returns $\mathcal{S}$ including $\{\langle p_2,p_5,p_8\rangle, \langle  p_2, p_5, p_7\rangle \}$.
The length score of $\langle p_2, p_5, p_7\rangle$ is 12, which is less than the length score of $\langle p_2,p_5,p_8\rangle$ of 15.
\end{example}

\subsubsection{Tightening upper bound: Arranging routes in the priority queue} 
\label{sec:priority_queue}

We use the upper bound to prune unnecessary routes. 
The upper bound  is computed from the obtained sequenced routes.
To tighten the upper bound, it is important to efficiently find sequenced routes that have small length and semantic scores.
{\sf BSSR} extends a route at the top of the priority queue to search for a sequenced route, as shown in Algorithm \ref{alg:bulk_ssr}.
Note that priority queues in existing algorithms conventionally consider only distances (i.e., a distance-based priority queue).
If we use a distance-based priority queue, {\sf BSSR} preferentially extends a route with a small length score.
Although we must increase the size of a route to $|\mathbf{S}_q|$ to find a sequenced route, a route that has a small length score likely has a small size.
Therefore, it is difficult to search for sequenced routes efficiently using a distance-based priority queue.

To search for sequenced routes efficiently, we preferentially extend a route that has a large size.
Here, since many routes in the priority queue could have the same size, we must consider an additional priority, which is expected to affect performance. 
If multiple routes in the priority queue are the same size, we preferentially extend the route with the smallest semantic score.
We can reduce the search space by searching for sequenced routes in ascending order of semantic score.
Moreover, if routes are the same size and have the same semantic score, we preferentially extend the route with the smallest length score.
As a result, we can efficiently obtain sequenced routes with small length and semantic scores.

\subsubsection{Tightening lower bound: Possible minimum length score}
\label{sec:lower}
As described in Section \ref{sec:branch}, we use the length scores of routes as the lower bound, i.e., we prune a route if the length score of the route is not less than the threshold.
Note that the length score of the route increases as the route size increases.
This indicates that it is difficult to prune routes before the route size increases. 
Our approach to tighten the lower bound of the route is to estimate the increase of the length score.
However, if we carelessly estimate a future length score, we may sacrifice the exactness of th result.

The basic idea of this estimation is to calculate the {\it possible minimum distance}. Here, we compute the smallest distance among any pair of PoI vertices in sets of PoI vertices.
We use the following two minimum distances, {\it semantic-match minimum distance} $\underline{l_s}$ and {\it perfect-match minimum distance} $\underline{l_p}$:

\begin{definition}({\bf minimum distance})\label{def:irrelevant}
The semantic-match minimum distance $\underline{l_s}$ and perfect-match minimum distance $\underline{l_p}$  are given by the following equations:
\begin{equation}\label{eq:irrelevant}
\underline{l_s}(\mathbf{R})\!=\! \Sigma_{i=|\mathbf{R}|}^{|\mathbf{S}_q|-1} \underline{l_s}[i], {\rm where}~ \underline{l_s}[i] \!=\!\!\!\!\! \min_{p_i\! \in \mathbb{P}_{t_i}, p_{i\!+\!1} \!\in \mathbb{P}_{t_{i\!+\!1}}} \!\!\!\!\!\!\!\!\!D(p_i, p_{i\!+\!1}).
\end{equation}
\begin{equation}\label{eq:zerosemantic}
\underline{l_p}(\mathbf{R})\!=\! \Sigma_{i=|\mathbf{R}|}^{|\mathbf{S}_q|-1} \underline{l_p}[i], {\rm where}~ \underline{l_p}[i]\! =\!\!\!\!\!\min_{p_i \!\in \mathbb{P}_{t_i}, p_{i\!+\!1} \!\in \mathbb{P}_{c_{i\!+\!1}}} \!\!\!\!\!\!\!\!\!D(p_i, p_{i\!+\!1}).
\end{equation}
In Equations  (\ref{eq:irrelevant}) and (\ref{eq:zerosemantic}), $\mathbb{P}_{t_i}$ and $\mathbb{P}_{c_i}$ denote the set of PoI vertices associated with a category tree of $c_{S_q}[i]$ and the set of PoI vertices whose category is $c_{S_q}[i]$, respectively.
\end{definition}
\noindent
We compute the semantic-match minimum distance based on the distance  to the PoI vertices that semantically match the next category.
We can safely add the semantic-match minimum distance to the current length score without restriction. However, the semantic-match minimum distance is much less than the threshold.
Thus, it could be difficult to improve pruning performance; thus, we use the perfect-match minimum distance to increase pruning performance.
The perfect-match minimum distance is computed based on the distance to the PoI vertices that perfectly match the next category.
We can improve pruning performance using the perfect-match minimum distance compared to the semantic-match minimum distance because the perfect-match minimum distance is much greater than the semantic-match minimum distance; therefore, the perfect-match minimum distance tightens the lower bound more than the semantic-match minimum distance. 
However, we can use the perfect-match minimum distance only in a special case, i.e., where a route must pass only PoIs that perfectly match the given categories  so as not to be dominated.
The perfect-match minimum distance works well if the number of sequenced route in $\mathcal{S}$ is large because the constraint is usually satisfied by increasing the number of sequenced route in $\mathcal{S}$. 

\begin{lemma}\label{lemma:zerosemantic}
Let $\mathbf{R}'$ and $\mathbf{R}''$ be sequenced routes in $\mathcal{S}$ and $\mathbf{R}$ be a route such that (i) $l(\mathbf{R}) \geq l(\mathbf{R}')$ and $s(\mathbf{R}) < s(\mathbf{R}')$ and (ii) $l(\mathbf{R}) < l(\mathbf{R}'')$ and $s(\mathbf{R}) \geq s(\mathbf{R}'')$.
Let $\delta$ be the minimum increment of a semantic score\footnote{The least increase of the semantic score is computed from the category tree. Specifically, we can compute the least increase from the category that is most similar (but not equal) to the next category.}.
We can prune $\mathbf{R}$ if we have (a) $l(\mathbf{R}) \geq l(\mathbf{R}')$ and $s(\mathbf{R}) + \delta \geq s(\mathbf{R}')$ and (b) $l(\mathbf{R})+\underline{l_p}(\mathbf{R}) \geq l(\mathbf{R}'')$ and $s(\mathbf{R}) \geq s(\mathbf{R}'')$.
\end{lemma}
{\it proof:} First, we consider case (a). If we have $l(\mathbf{R}) \geq l(\mathbf{R}')$ and $s(\mathbf{R}) + \delta \geq s(\mathbf{R}')$, $\mathbf{R}$ is dominated by or equivalent to $\mathbf{R}'$ if its semantic score increases. 
Therefore, $\mathbf{R}$ must only pass through PoI vertices that perfectly match the given categories not to be dominated.
If $\mathbf{R}$ passes through only PoI vertices that perfectly match the given categories, the length score of $\mathbf{R}$ increases by at least $\underline{l_p}(\mathbf{R})$.
For case (b), if we have $l(\mathbf{R})+\underline{l_p}(\mathbf{R}) \geq l(\mathbf{R}'')$ and $s(\mathbf{R}) \geq s(\mathbf{R}'')$, $\mathbf{R}$ is dominated by or equivalent to $\mathbf{R}''$ if its length score increases by $\underline{l_p}(\mathbf{R})$.
As a result, if we have two routes $\mathbf{R}'$ and $\mathbf{R}''$, such as (i) $l(\mathbf{R}) \geq l(\mathbf{R}')$ and $s(\mathbf{R}) + \delta \geq s(\mathbf{R}')$ and (ii) $l(\mathbf{R})+\underline{l_p}(\mathbf{R}) \geq l(\mathbf{R}'')$ and $s(\mathbf{R}) \geq s(\mathbf{R}'')$, $\mathbf{R}$ is dominated by or equivalent to at least one of $\mathbf{R}'$ and $\mathbf{R}''$. \hfill{}$\square$

To compute the estimation of the lower bound, we compute two types of possible minimum distances $\underline{l_s}$ and $\underline{l_p}$.
A naive approach computes all minimum distances from the PoI vertices that semantically match $c_{S_q}[i]$ to $c_{S_q}[i+1]$ for $1 \leq i \leq |\mathbf{S}_q|-1$ by iteratively executing the Dijkstra algorithm.
However, this has a high computational cost.
To reduce the cost, we execute a {\it multi-source multi-destination Dijkstra algorithm}.
In this algorithm, all start points are inserted into the same priority queue.
Then, the algorithm dequeues vertices in the same manner as the conventional Dijkstra algorithm.
Here, the process is terminated if the top of the priority queue becomes one of the destinations.
This approach only needs $|\mathbf{S}_q|-1$ times to compute the possible minimum distance. The multi-source multi-destination Dijkstra algorithm guarantees the minimum distance by the following lemma:

\begin{lemma}{}\label{lemma:multidikstra}
The multi-source multi-destination Dijkstra algorithm guarantees the minimum distance from the start points to the destinations.
\end{lemma}
{\it proof:} We first insert multiple start points into the priority queue, and their distances from the start points are initialized as 0. If we find a vertex, it is inserted into the queue and the distance to the vertex is updated from the closest start point to the vertex. The vertex with the smallest distance from the start point in the priority queue is dequeued from the priority queue. If the top vertex in the priority queue is one of the destinations, there are no destinations with smaller distance than the top one. Therefore, we can guarantee the minimum distance from the start points to the destinations. \hfill{}$\square$

Algorithm \ref{alg:lowerbound} shows the pseudocode to compute the semantic-match minimum distance.
The estimation of the lower bound is executed after line 4 in Algorithm \ref{alg:bulk_ssr}.
Here, we initialize $\mathbb{P}_i$ and $\mathbb{P}_{i+1}$ (lines 3--4).
$\overline{l}(\phi)$ denotes the threshold for a route whose semantic score is 0.
The difference between computing the semantic-match and perfect-match minimum distances is whether the PoI vertices in $\mathbb{P}_{i+1}$ semantically or perfectly match the given category.

\begin{algorithm}[h]
{\footnotesize
	\caption{Computing possible minimum distance}	\label{alg:lowerbound}
	            \SetKwFunction{Estimation}{EstimationLowerbound}
	            \SetKwInOut{Input}{input}\SetKwInOut{Output}{output}
	            {\bf procedure} \Estimation{$v_q$, $\mathbf{S}_q$}\\
		              \For{ $i$ : 1 to $|\mathbf{S}_q|-1$ }{
				               $\mathbb{P}_i \leftarrow \{ p | p \in \mathbb{P}_{t_{c_{S_q}[i]}}$ and $D(v_q, p) < \overline{l}(\phi) \}$;\\			     
				               $\mathbb{P}_{i+1} \leftarrow \{ p | p \in \mathbb{P}_{t_{c_{S_q}[i+1]}}$ and $D(v_q, p) < \overline{l}(\phi) \}$;\\				               				              
				              $dist[u] = \inf$ for all $u \in \mathbb{V} \cup \mathbb{P}$, $dist[p]=0$ for all $p \in \mathbb{P}_i$;\\       
				              priority$\_$queue $Q$ $\leftarrow$ $\{p \} \in \mathbb{P}_i$;\\
					          \While{$Q$ is not empty}{
						          $u$ $\leftarrow$ $Q$.dequeue;\\
						          \lIf{$u$ is already visited}{continue}				          					         
						          \If {$u \in \mathbb{P}_{i+1}$}{
							          $\underline{l_s}[i]=dist[u]$; \\
							         break;
						          }
						          \For{each $u'$ for $e(u,u') \in E$}{
							         \If{$dist[u] +  w(u,u') < dist[u']$ }{
								          $dist[u']=dist[u] +  w(u,u')$;\\
								          $Q$.enqueue($u'$); \\
								          }
						         }
				            }
	               }
	               {\bf return} $\underline{l_s}$;\\
	               {\bf end procedure}
}
\end{algorithm}

\begin{example}
We show an example to compute the semantic-match minimum distance using Example \ref{ex:intro}.
$\mathbb{P}_1$, $\mathbb{P}_2$, and $\mathbb{P}_3$ include  $\{p_1,p_2,p_6,p_{10},p_{11}\}$,  $\{p_5,p_{9},p_{12}\}$, and $\{p_3,p_{4},p_{7},p_{8},p_{13}\}$, respectively.
First, PoI vertices in $\mathbb{P}_1$ are inserted to priority queue $Q$, and the set of destinations is $\mathbb{P}_2$.
By processing the Dijkstra algorithm, we compute possible minimum distance $\underline{l_s}[1] = 2 $ (from $p_6$ to $p_9$).
Next, we search PoI vertices that semantically match A$\&$E to gift shop.
Then, we compute $\underline{l_s}[2] =1$ (from $p_{12}$ to $p_{13}$).
Finally, we obtain semantic-match minimum distance $\underline{l_s} = \{2,1\}$.
We can compute the perfect-match minimum distance in the same way and obtain $\underline{l_p}=\{3,1\}$, which is greater than $\underline{l_s}$. 
\end{example}

\subsubsection{Reuse of the temporal result: On-the-fly caching technique} 
\label{sec:cache}
Although {\sf BSSR} efficiently prunes unnecessary routes, it may iteratively execute the modified Dijkstra algorithm at the same vertex because, in Algorithm \ref{alg:bulk_ssr} (line 8), $p_d$ could be the same as the former executions of the modified Dijkstra algorithms.
Thus, we reuse the result starting at the same PoI vertex by materializing the result of the modified Dijkstra algorithm (i.e., keeping PoI vertices matching $c_d$ and distances from $p_d$ to the PoI vertices), which we refer to as {\it on-the-fly caching}.

After finding SkySRs, on-the-fly caching frees the results of the modified Dijkstra algorithms (this is why we call it {\it on-the-fly}), because the search space rarely overlaps across different inputs (i.e., $\mathbf{S}_q$ and $v_q$ differ).

\subsection{Theoretical Analysis}
\label{sec:cost}
In this section, we theoretically analyze the cost and correctness of the proposed {\sf BSSR}.

\begin{theo}{{\bf (Time complexity)}}
Let $\gamma$ be a ratio of pruning and $\alpha$ be a ratio of the size of a graph to find the SkySRs.
The time complexity of  {\sf BSSR} is $O(\gamma (\alpha |\mathbb{P}|)^{|\mathbf{S}_q|}\alpha(|\mathbb{E}|+(|\mathbb{V}| + |\mathbb{P}|) \log (\alpha (|\mathbb{V}| +  |\mathbb{P}|))))$.
\end{theo}
{\it proof:} The time complexity of the Dijkstra algorithm is  $O(|\mathbb{E}|+|\mathbb{V}| \log |\mathbb{V}|)$ if the number of vertices is $|\mathbb{V}|$. 
In our setting, we have $|\mathbb{V}| + |\mathbb{P}|$ vertices because we have two types of vertices.
In addition, we do not need to search the whole graph by reducing the graph size according to the threshold.
Therefore, the time complexity of the modified Dijkstra algorithm is $O(\alpha (|\mathbb{E}|+(|\mathbb{V}| + |\mathbb{P}|) \log (\alpha (|\mathbb{V}| +  |\mathbb{P}|)))$.
The time complexity of {\sf BSSR} depends on the number of times the modified Dijkstra algorithms is executed.
The number of modified Dijkstra algorithms is equal to all the potential routes $|\mathbb{P}|^{|\mathbf{S}_q|}$. Recall that we can prune the number of routes using the branch-and-bound algorithm. Finally, the time complexity of  {\sf BSSR} is $O(\gamma (\alpha |\mathbb{P}|)^{|\mathbf{S}_q|}\alpha( |\mathbb{E}|+(|\mathbb{V}| + |\mathbb{P}|) \log (\alpha(|\mathbb{V}| + |\mathbb{P}|))))$. \hfill{} $\square$

In our approach, $\gamma$ and $\alpha$ depend  on the upper and lower bounds.
These are  affected  by the graph structure, the category trees, and the ratio of PoI vertices, and the time complexity of {\sf BSSR} depends on these factors.

\begin{theo}{{\bf (Space complexity)}}
Let $\gamma$ be the pruning ratio, and $\alpha$ be the ratio of the size of the graph to find the SkySRs.
The space complexity of {\sf BSSR} is $O(|\mathbb{E}|+|\mathbb{V}|+|\mathbb{P}|+\gamma |\mathbf{S}_q| (\alpha |\mathbb{P}|)^{|\mathbf{S}_q|})$.
\end{theo}
{\it proof:} We store the whole graph of size $O(|\mathbb{E}|+|\mathbb{V}|+|\mathbb{P}|)$.
We also store routes into the priority queue and $\mathcal{S}$, and the maximum number of routes is $|\mathbb{P}|^{|\mathbf{S}_q|}$.
We can prune the number of routes using the branch-and-bound algorithm. 
The size of the routes is proportional to $|\mathbf{S}_q|$. Therefore, the space complexity of {\sf BSSR} is $O(|\mathbb{E}|+|\mathbb{V}|+|\mathbb{P}|+\gamma |\mathbf{S}_q| (\alpha|\mathbb{P}|)^{|\mathbf{S}_q|})$. \hfill{}$\square$

If the number of routes in the priority queue is small, the graph size becomes the main factor related to the memory usage.
Otherwise, the number of routes in the priority queue is the main factor.

\begin{theo}{{\bf (Correctness)}}
{\sf BSSR} guarantees the exact result. 
\end{theo}
{\it proof:} {\sf BSSR} prunes routes based on the upper and lower bounds. {\sf BSSR} safely prunes routes dominated by or equivalent to the obtained sequenced routes. 
As a result,  {\sf BSSR} does not sacrifice the exactness of the search result. \hfill{} $\square$

 \subsection{Running Example}
 \label{sec:runex}
 
 We demonstrate {\sf BSSR} with optimization techniques using Example \ref{ex:intro}.
 Table \ref{table:BSSR} shows routes in priority queue $Q_b$ and sequenced routes in $\mathcal{S}$.
 To compute category similarity and semantic score, we use Equations (6) and (7), respectively.
 
 First, we process {\sf NNinit}, and $\mathcal{S}$ initially includes $\{\langle p_2,p_5,p_8\rangle$, $\langle p_2, p_5, p_7\rangle \}$.
 1st step: {\sf BSSR} starts to find PoI vertices that semantically match Asian restaurant from $v_q$ with the threshold of 15. Then, it finds $p_1$, $p_2$, $p_6$, $p_{10}$, and $p_{11}$.
 Both $p_2$'s  and $p_{10}$'s category similarities are 1, and their lengths are 6 and 8, respectively. Thus, $p_2$ comes the top in $Q_b$.
 2nd step: {\sf BSSR} searches PoI vertices that semantically match Arts$\&$Entertainment from $p_2$, and finds $p_5$.
 Since $\langle p_2, p_{12}\rangle$ passes through $p_5$ and $l(\langle p_2, p_9 \rangle)$ is more than 15, both routes are not inserted to $Q_b$.
 3rd step: as the top route is $\langle p_2, p_5 \rangle$, {\sf BSSR} searches PoI vertices that semantically match gift shop from $p_5$.
 {\sf BSSR} does not find any routes due to the threshold.
 4th step: {\sf BSSR} fetches $\langle p_{10}\rangle$ from $Q_b$ and inserts two routes $\langle p_{10},p_{5}\rangle$ and $\langle p_{10},p_{12}\rangle$ to $Q_b$.
 5th step: {\sf BSSR} fetches  $\langle p_{10},p_{12}\rangle$ and finds sequenced route $\langle p_{10},p_{12}, p_{13}\rangle$.
 Since $\langle p_{10},p_{12}, p_{13}\rangle$ dominates $\langle p_2, p_5, p_8\rangle$,  $\langle p_2, p_5, p_8\rangle$  is deleted from $\mathcal{S}$.
 6th step: The top route $\langle p_{10}, p_{5}\rangle$ is deleted from $Q_b$ because its length score is not smaller than the threshold of 13.
 7th step: {\sf BSSR} fetches $\langle p_1 \rangle$ and inserts $\langle p_1, p_5\rangle$ and $\langle p_1, p_9\rangle$.
 8th step: {\sf BSSR} fetches $\langle p_1, p_9\rangle$ and finds a sequenced route $\langle p_1, p_9, p_8 \rangle$.
  $\langle p_1, p_9, p_8 \rangle$ is inserted to $\mathcal{S}$, and $\langle p_2, p_5, p_7\rangle$  is deleted from $\mathcal{S}$.
 9th step: $\langle p_1, p_5 \rangle$ is deleted due to the threshold.
 10th step: {\sf BSSR} fetches $\langle p_6\rangle$ and finds a route $\langle p_6, p_9 \rangle$. 
 11th step: {\sf BSSR} finds a sequenced route  $\langle p_6,p_9, p_8 \rangle$, and the route dominates $\langle p_1,p_9,p_8 \rangle$.
 12th step: The distance from $p_{11}$ to the PoI vertices that match A$\&$E is larger than the threshold. Finally, {\sf BSSR} returns the set of SkySRs $\mathcal{S}$.
 
 \begin{table}[t]
 	\begin{center}
 	\caption{Example of BSSR algorithm}	\label{table:BSSR}
 			\scriptsize
 		\begin{tabular}{|l|l|} \hline
 0&$Q_b$: \\
 & $\mathcal{S}$: $\langle p_2,p_5,p_8\rangle, \langle p_2, p_5, p_7\rangle$\\\hline
 1& $Q_b$: $\langle p_2 \rangle, \langle p_{10} \rangle , \langle p_1 \rangle , \langle p_6 \rangle, \langle p_{11} \rangle $\\
 & $\mathcal{S}$: $\langle p_2,p_5,p_8\rangle, \langle p_2, p_5, p_7\rangle$\\ \hline
 2& $Q_b$: $\langle p_2,p_5\rangle,\langle p_{10}\rangle,\langle p_1\rangle, \langle p_6\rangle, \langle p_{11}\rangle $\\
 & $\mathcal{S}$: $\langle p_2,p_5,p_8\rangle, \langle p_2, p_5, p_7\rangle$\\ \hline
 3& $Q_b$: $\langle p_{10}\rangle ,\langle p_1\rangle, \langle p_6\rangle, \langle p_{11}\rangle$\\
 &$\mathcal{S}$: $\langle p_2, p_5, p_8\rangle ,\langle p_2,p_5,p_7\rangle$\\ \hline
 4& $Q_b$: $\langle p_{10}, p_{12}\rangle ,\langle p_{10}, p_{5}\rangle ,\langle p_1\rangle , \langle p_6\rangle, \langle p_{11}\rangle$\\
 &$\mathcal{S}$: $\langle p_2, p_5, p_8\rangle ,\langle p_2,p_5,p_7\rangle$\\ \hline
 5& $Q_b$: $\langle p_{10}, p_{5}\rangle ,\langle p_1\rangle, \langle p_6\rangle, \langle p_{11}\rangle$\\
 &$\mathcal{S}$: $\langle p_{10}, p_{12},p_{13}\rangle ,\langle p_2,p_5,p_7\rangle$\\ \hline
 6& $Q_b$: $\langle p_1\rangle, \langle p_6\rangle, \langle p_{11}\rangle$\\
 &$\mathcal{S}$: $\langle p_{10}, p_{12},p_{13}\rangle,\langle p_2,p_5,p_7 \rangle$\\ \hline
 7& $Q_b$: $\langle p_1, p_9\rangle, \langle p_1, p_5\rangle, \langle p_6\rangle , \langle p_{11} \rangle$\\
 &$\mathcal{S}$: $\langle p_{10}, p_{12},p_{13}\rangle,\langle p_2,p_5,p_7 \rangle$\\ \hline
 8& $Q_b$: $\langle p_1, p_5 \rangle, \langle p_6 \rangle, \langle p_{11}\rangle$\\
 &$\mathcal{S}$: $\langle p_{10}, p_{12},p_{13}\rangle ,\langle p_1, p_9, p_8\rangle$\\ \hline
 9& $Q_b$: $\langle p_6\rangle , \langle p_{11} \rangle$\\
 &$\mathcal{S}$: $\langle p_{10}, p_{12},p_{13}\rangle , \langle p_1, p_9, p_8 \rangle$\\ \hline
 10& $Q_b$:  $\langle p_6,p_9 \rangle , \langle p_{11}\rangle $\\
 &$\mathcal{S}$: $\langle p_{10}, p_{12},p_{13} \rangle , \langle p_1, p_9, p_8 \rangle $\\ \hline
 11& $Q_b$:  $\langle p_{11}\rangle$\\
 &$\mathcal{S}$: $\langle p_{10}, p_{12},p_{13}\rangle , \langle p_6,p_9, p_8 \rangle$ \\ \hline
 12&$Q_b$:\\
 &$\mathcal{S}$: $\langle p_{10}, p_{12},p_{13}\rangle , \langle p_6,p_9, p_8\rangle$\\ \hline
 \end{tabular}
 \end{center}
 \end{table}

\section{Variations  and extensions}
\label{sec:dis}
The SkySR query has a number of variations and extensions. We discuss some of these in the following.

\noindent
{\bf Directed graphs:} The SkySR query can be easily applied to directed graphs.
We only need to use the Dijkstra algorithm for directed graphs. Here, no modification of the main idea is required.

\noindent
{\bf PoI with multiple categories:} To treat PoIs with multiple categories, we can change the definitions of sequenced routes and category similarity.
Specifically, we change condition (ii) in Definition \ref{def:seqencedroute} to state that at least one $c_{p_i}[j]$ $(1\leq j \leq k_i)$ semantically matches $c_S[i]$ for $1 \leq i \leq |\mathbf{S}|$, where $c_{p_i}[j]$ is the $j$-th category of $p_i$ and $k_i$ is the number of categories associated with $p_i$.
The category similarity is either the highest or the average value among the category similarities.

\noindent
{\bf Complex category requirement:} We can specify more detailed category requirements, such as {\it conjunction}, {\it disjunction}, and {\it negation}.
For example, we can specify that a PoI category is  ``American restaurant'' or ``Mexican restaurant'' (disjunction), but not ``Taco Place'' (negation).
If PoI vertices are associated with more than two categories, we can specify a  conjunction such as ``Cafe'' and ``Bakery''.
Note that the time complexity of our algorithm does not change if we specify a detailed requirement because the detailed requirements are equivalent to increasing the number of categories.

\noindent
{\bf Skyline trip planning query:} The proposed algorithm can be applied to the trip planning query without category order.
For searching routes without category order, the proposed algorithm searches PoI vertices that semantically match a category in a given set of categories. 
Then, if the algorithm finds PoI vertices, it deletes the categories that are already included in the routes to find next PoI vertices. 
Note that we need to modify some definition and scoring functions for routes without category order.
By this procedure, we can find skyline routes efficiently.

\noindent
{\bf SkySR with destination:} Note that we can specify the destination.
The simple way to calculate a SkySR with a destination is to add the distance from the last visited PoI vertex to the destination to the length score after finding the sequenced route.
To improve efficiency, we traverse PoI vertices from both the destination and the start point.

\section{Experimental study}
\label{sec:exp}
We perform experiments to evaluate the effectiveness of the proposed algorithm.
All algorithms are implemented in C++ and run on an Intel(R) Xeon(R) CPU E5620 @ 2.40GHz with 32 GB of RAM.

\begin{figure*}[ttt]
\centering
 \subfloat[Tokyo]{\epsfig{file=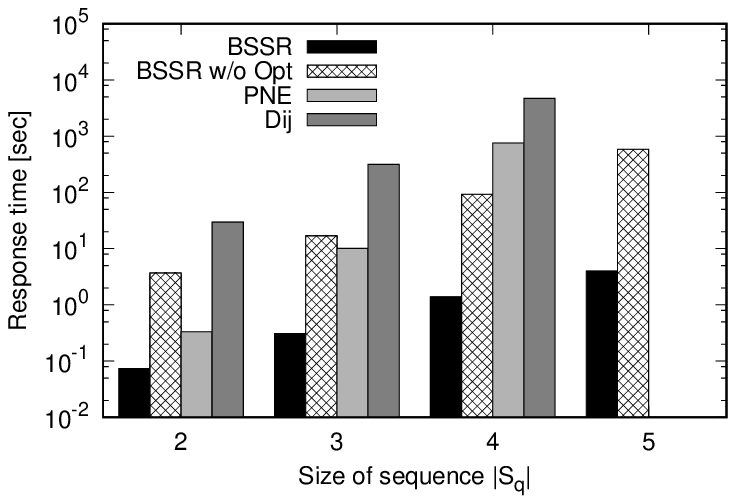,width=0.33\linewidth}
 \label{fig:cpu1}}
 \subfloat[NYC]{\epsfig{file=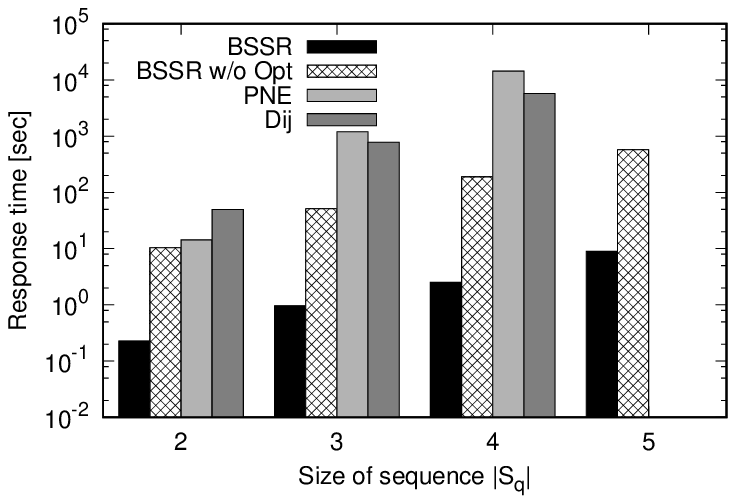,width=0.33\linewidth}
 \label{fig:cpu1}}
  \subfloat[Cal]{\epsfig{file=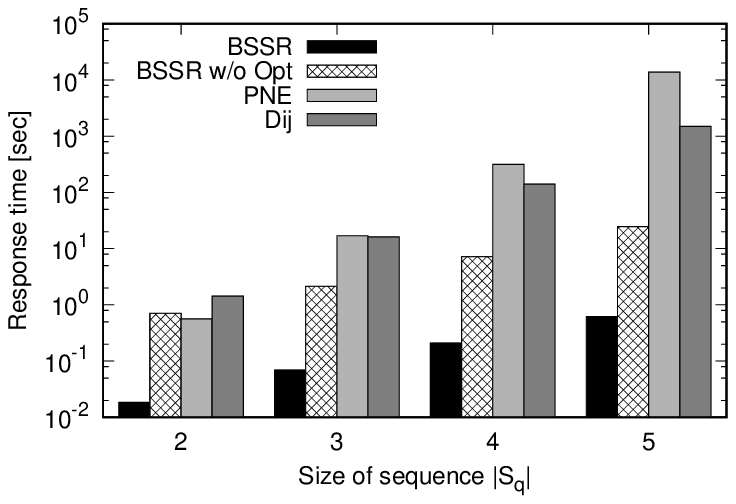,width=0.33\linewidth}
  \label{fig:cpu1}}
  \vspace{-1mm}
 \caption{Results obtained for the datasets with various $|\mathbf{S}_q|$}
 \label{fig:overview}
 \end{figure*}

\subsection{Experimental settings}
\noindent
{\bf Algorithm.}
We compare the proposed {\sf BSSR} and algorithms that iteratively find OSRs using the Dijkstra-based solution and the PNE approach (denoted {\sf Dij} and {\sf PNE}, respectively), as described in Section \ref{sec:pre}.
We evaluate performance with respect to (i) response time, and (ii) maximum resident set size (RSS) to represent memory usage.

\begin{table}[t]
 	\begin{center}
	\caption{Summery of dataset} 
		\label{table:graphs}
	\footnotesize
		\begin{tabular}{|l|l|r|r|r|r|} \hline
\multicolumn{1}{|l|}{Dataset} & \multicolumn{1}{|l|}{Area} &  \multicolumn{1}{|l|}{$|\mathbb{V}|$} & \multicolumn{1}{|l|}{$|\mathbb{P}|$} & \multicolumn{1}{|l|}{$|\mathbb{E}|$} \\\hline
Tokyo &  Tokyo & 401,893& 174,421		& 499,397	 	 		\\
NYC & New York city & 1,150,744& 451,051 	& 1,722,350  \\
Cal & California & 21,048& 87,365	& 108,863	 	 		 \\
\hline
\end{tabular}
\end{center}
\end{table}

\noindent
{\bf Dataset.}
We conduct experiments using various maps (Tokyo, New York city, and California).
Table \ref{table:graphs} summarizes each dataset.
For the Tokyo and NYC datasets, the road network is extracted from OpenStreetMap\footnote{https://www.openstreetmap.org} and the PoI information is extracted from Foursquare.
Each PoI is embedded on the closest edge in the same way as \cite{li2005trip} and is associated with the Foursquare category trees. Note that the number of category trees in Foursquare is 10.
For the Cal dataset, the road network and PoI information are available online\footnote{http://www.cs.utah.edu/$\sim$lifeifei/SpatialDataset.htm}. The number of categories in the Cal dataset is 63\footnote{Since the PoIs in the Cal dataset have no category tree information, we generate a category of height three where a non-leaf node has three child nodes.}.
For each dataset, we use distances based on longitude and latitude as edge weights and treat the graphs as undirected graphs. 
The graphs are implemented using adjacency lists.

For each dataset, we generate 100 searches, in which the size of a sequence is $|\mathbf{S}_q|$.
The start points are selected randomly from vertices in the maps.
The categories of sequences are selected randomly from the leaf nodes in the category trees with the constraint that they have different category trees.
Since the number of PoI vertices associated with each category is significantly biased, we select only categories that have a large number of PoI vertices.

Here, category similarity is calculated based on the \emph{Wu and Palmer similarity measure} \cite{wu1994verbs} and the semantic score is calculated as the product of the category similarities of the sequence members. Specifically, we calculate the category similarity and semantic score using the following equations:
\begin{eqnarray} 
 sim(c,c') &=& {\rm max}_{c_i \in a(c')} {\textstyle \frac{2 \cdot d(c_m)}{d(c) + d(c')}}, \\
 s(\mathbf{R})&=&  1 - \Pi_{i=1}^{\min (|\mathbf{R}|,|\mathbf{S}_q|)} sim(c_{p_{R}[i]}, c_{S_q}[i]),
\end{eqnarray}
\noindent
where $a(c)$, $d(c)$, and $c_m$ denote the set of ancestor categories of $c$ (including $c$), the depth of $c$,  and the deepest  common ancestor category of $c$ and $c_i$, respectively.

\subsection{Overview of results}

First, we present an overview of the performance of all algorithms.
Figure \ref{fig:overview} shows the response time with various category sequence sizes, and Table \ref{table:overview} shows the RSS for a category sequence of size four. Here, ``{\sf BSSR w/o Opt}'' denotes {\sf BSSR} without optimization techniques. In Figure \ref{fig:overview}, there are missing bars for the case of size of sequence 5, because the executions were not finished after a month.

{\sf BSSR} achieves the least response time with all datasets and reduces the search space by exploiting the branch-and-bound algorithm and the proposed optimization techniques.
By comparing {\sf BSSR} and {\sf BSSR w/o Opt}, we confirm that the optimization techniques increase efficiency.
When the size of the category sequence is small, {\sf PNE} finds SkySRs efficiently because it can search for sequenced routes efficiently if the category sequence size is small.
 On the other hand, as category sequence size increases, the response time of {\sf PNE} and {\sf Dij} increases significantly.
If the category sequence size is large, {\sf BSSR} achieves better performance than {\sf PNE} and {\sf Dij} even if we do not use optimization techniques.
By comparing {\sf Dij} to {\sf PNE}, it can be seen that their performance depends on the datasets and the category sequence size.
Although the PNE approach was proposed to be a more sophisticated algorithm than the Dijkstra-based solution \cite{sharifzadeh2008optimal}, {\sf PNE} requires more time than {\sf Dij} for the NYC and Cal datasets, which
implies that it is not effectively robust to datasets.
In terms of RSS, {\sf BSSR} and {\sf PNE} achieve nearly the same performance.
These two algorithms do not store many routes in the priority queue; therefore, RSS is highly dependent on the graph size.
On the other hand, as {\sf Dij} stores many routes in the priority queue, RSS is significantly larger than those of the other algorithms.
Although we do not show the routes returned by each algorithm due to space limitations, all algorithms output the same routes.  
As a result, {\sf BSSR} achieves the fastest response time with small memory usage without sacrificing the exactness of the result.

\begin{figure*}[ttt]
  \begin{minipage}[]{.23\textwidth}
 	\centering
 \epsfig{file=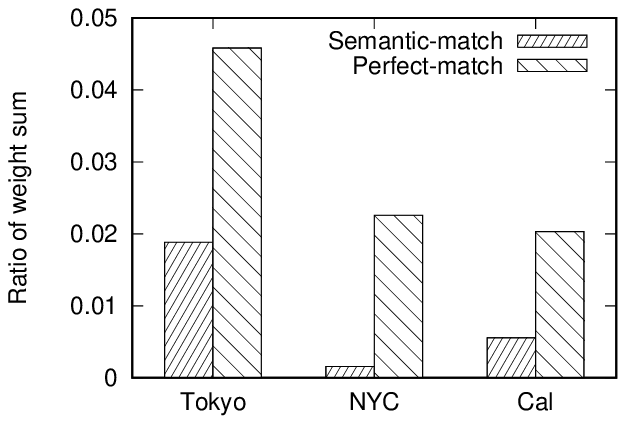,width=1.0\linewidth}
 \caption{Effect of minimum possible distances}
  \label{fig:lowerbound}
  \end{minipage}
  \hfil
   \begin{minipage}[]{.75\textwidth}
   \centering
    \subfloat[Tokyo]{\epsfig{file=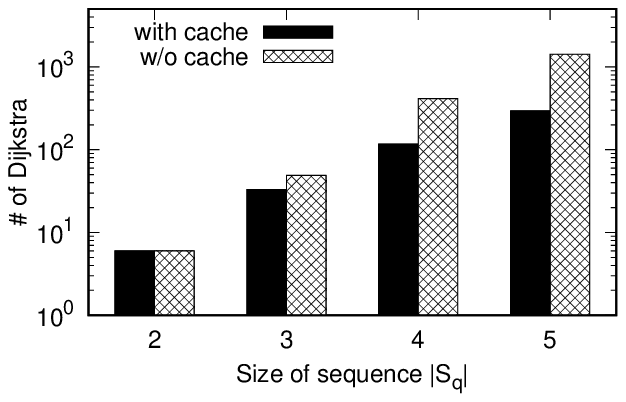,width=0.3\linewidth}
    \label{fig:cpu1}}
    \subfloat[NYC]{ \epsfig{file=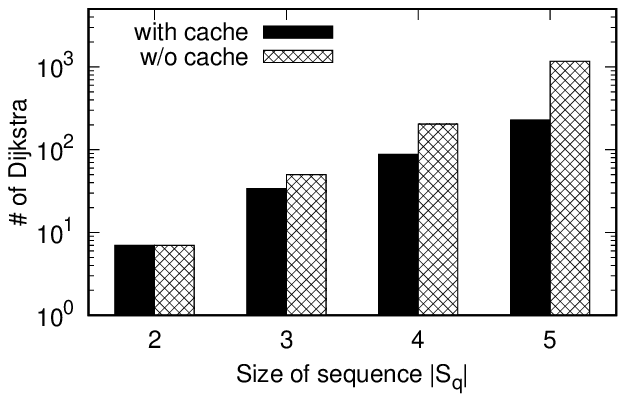,width=0.3\linewidth}
    \label{fig:cpu1}}
	\vspace{-1mm}
	 \subfloat[Cal]{ \epsfig{file=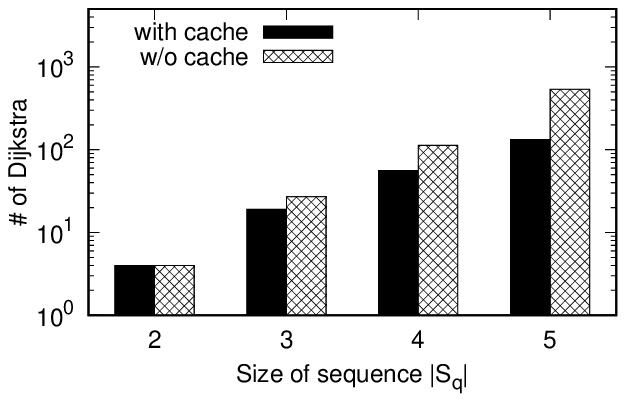,width=0.3\linewidth}
	\label{fig:cpu1}}
	\vspace{-1mm}
    \caption{Effect of on-the-fly caching for various $|\mathbf{S}_q|$}
    \label{fig:cache}
 \end{minipage}
 \end{figure*}

\begin{table}[t]
 	\begin{center}
 	\caption{RSS Comparison}	
 	\label{table:overview}
 	\footnotesize
 		\begin{tabular}{|l|r|r|r|r|} \hline
 		               &\multicolumn{1}{|l|}{{\sf BSSR}}&\multicolumn{1}{|l|}{{\sf BSSR w/o Opt}}&\multicolumn{1}{|l|}{{\sf PNE}}&	\multicolumn{1}{|l|}{{\sf Dij}}  \\\hline
 		Tokyo   	& 239.6 MB &	497.5 MB 	& 239.8 MB	&	 4.8 GB						\\
 		NYC 		 	& 658.0 MB  &  659.4 MB 	& 658.7 MB	&	 9.7 GB			\\	
 		Cal	  		&	36.7 MB	&	53.7 MB	  	&	36.6 MB	& 70.3 MB \\\hline		
 \end{tabular}
 \end{center}
 \end{table}
 
\begin{table}[t]
	\begin{center}
	\caption{Effect of initial search for various $|\mathbf{S}_q|$} 
		\label{table:initial}
		\scriptsize
		\begin{tabular}{|l|l|l|r|r|r|r|} \hline
Dataset&Approach& Metrics  &2&3&4&5 \\\hline
\multirow{5}{*}{Tokyo} &\multirow{4}{*}{Proposed} &Weight sum&	0.009&	0.013&	0.017&	0.021 \\	             
 &&Response time [msec]	&3.5&	5.1&	6.9&	8.6 \\
 &&$\#$ of routes&	1.49&	1.33&	1.36&	1.49 \\
 &&Ratio&	0.74&	0.79&	0.82	& 0.86 \\\cline{2-7}
&Existing & Weight sum &  \multicolumn{4}{|c|}{0.32 (regardless $|\mathbf{S}_q|$)} \\
\hline
\multirow{5}{*}{NYC} &\multirow{4}{*}{Proposed} &Weight sum&	0.044&	0.066&	0.073&	0.078 \\	             
 &&Response time [msec]	&10.7&	16.5&	19.5&	24.1 \\
&&$\#$ of routes&	1.76&	1.79&	1.81&	1.82\\
&&Ratio&	0.67&	0.81&	0.85&	0.83\\\cline{2-7}
&Existing & Weight sum &  \multicolumn{4}{|c|}{1.31 (regardless $|\mathbf{S}_q|$)} \\
\hline		
\multirow{5}{*}{Cal} &\multirow{4}{*}{Proposed} & Weight sum & 0.79 & 1.28	& 1.57	& 1.85	 \\
&& Response time [msec] & 1.4&	2.3	& 2.9 & 3.9 \\
&&$\#$ of routes & 2.27&	2.37&	2.28	&2.25	\\
&& Ratio & 0.70&	0.79&	0.85&	0.86		\\\cline{2-7}
&Existing & Weight sum &  \multicolumn{4}{|c|}{12.14 (regardless $|\mathbf{S}_q|$)} \\

\hline	              
\end{tabular}
\end{center}
\end{table}

\subsection{Optimization Techniques}
The optimization techniques improve the efficiency of {\sf BSSR}.
Here, we evaluate each optimization technique.

{\bf Initial Search:}
We show the search spaces with and without an initial search for the first modified Dijkstra algorithm to evaluate the effect of the initial search.
Moreover, we evaluate {\sf NNinit} in terms of response time. 
Table \ref{table:initial} shows the weight sum, which represents the search space, the response time of {\sf NNinit}, and the number of sequenced routes  found by {\sf NNinit} for various category sequence sizes.
In addition, we show the ratio of the length score of the sequenced route with the largest semantic score among the sequenced routes found in the initial search to the length score of the sequenced route whose semantic score is  0 in the initial search. 
The weight sum with the initial search is significantly smaller than that without the initial search.
We can avoid traversing the whole graph using the initial search; thus, this can significantly reduce the search space of {\sf BSSR}.
Moreover, since the response time of {\sf NNinit} is significantly less than that of {\sf BSSR} (Figure \ref{fig:overview}), we confirm that {\sf NNinit} can reduce the search space efficiently.
Note that the number of sequenced routes found by the initial search is not large. 
On the other hand, the length score of the sequenced route with the largest semantic score is much smaller than that of  the sequenced route whose semantic score is 0.
As a result, {\sf NNinit} reduces the search space significantly without increasing total response time.

{\bf Tightening Upper Bound:}
The priority queue aims at efficiently tightening the upper bound to reduce the search space.
Here, we show the total number of vertices visited by {\sf BSSR}, which is highly related to the response time.
Table \ref{table:upperbound} shows the total number of vertices visited  by the proposed priority queue and distance-based priority queue for various category sequence sizes.
The number of vertices visited by the proposed priority queue is less than that of the distance-based priority queue.
In particular, as the size of the category sequences increases, the performance gap increases because, as the category sequence size increases, the distance-based priority queue cannot find sequenced routes efficiently.
Thus, the upper bound is rarely updated.
On the other hand, the proposed priority queue can update the upper bound efficiently because the route with the largest size is dequeued preferentially.
Thus, the proposed priority queue is more suitable than the distance-based approach for finding  SkySRs. 

\begin{table}[t]
	\begin{center}
	\caption{Effect of priority queue for various $|\mathbf{S}_q|$} 
		\label{table:upperbound}
		\scriptsize
		\begin{tabular}{|l|l|r|r|r|r|} \hline
	Dataset&Approach & 2 & 3 & 4 & 5 \\\hline
	\multirow{2}{*}{Tokyo}&	{\sf Proposed }	   	    & $3750$ & $17600$	& $112000$ & 	$397000$								\\
		&{\sf  Distance-based} 	& $3890$ & $23500$	&	$189000$ & $1760000$			\\	\hline
	\multirow{2}{*}{NYC}&	{\sf Proposed }	   	    & $13800$& $108000$ 	&  $172000$ & $637000$								\\
		&{\sf  Distance-based} 		& $14800$& $165000$	&	$444000$ & $1520000$			\\\hline		
		\multirow{2}{*}{Cal} &{\sf Proposed}	   	    & $4900$& $24800$ 	&  $84900$ & $383000$								\\
		&{\sf Distance-based} 	& $5300$& $34900$	&	$168000$ & $899000$			\\	
		\hline		
\end{tabular}
\end{center}
\end{table}

{\bf Tightening Lower Bound:}
To tighten the lower bound, we propose two types of possible minimum distances, i.e., semantic-match and perfect-match minimum distances.
If the minimum possible distance is large, we can prune routes even if the routes include a small number of PoI vertices.
Figure \ref{fig:lowerbound} shows the ratios of the possible minimum distances to the sum weights of the initial search when we set the category sequence size to five.
The semantic-match and perfect-match minimum distances in the Tokyo dataset effectively reduce the search space by tightening the lower bound.
However, different from the Tokyo dataset, the possible minimum distances in the NYC and Cal datasets are small. 
Since the PoI vertices in the two datasets are relatively concentrated in a small area, the possible minimum distances become small.
The effect of the possible minimum distances highly depends on the skews of locations of the PoI vertices.

{\bf On-the-fly Caching:}
On-the-fly caching can reuse the results of former modified Dijkstra algorithm executions; thus, the number of executions of the Dijkstra algorithm decreases.
Figure \ref{fig:cache} shows the numbers of executions of modified Dijkstra algorithms by {\sf BSSR} with all optimization techniques and those except for on-the-fly caching.
The number of executions of the Dijkstra algorithms decreases using on-the-fly caching.
In particular, when the category sequence size increases, the performance gap increases because, as the category sequence size increases, we have more opportunities to reuse former results.
Thus, we confirm that on-the-fly caching is effective to reduce the number of executions of the Dijkstra algorithms.

 \begin{figure}[t]
 	 	\centering
	 	\includegraphics[width=0.8\linewidth]{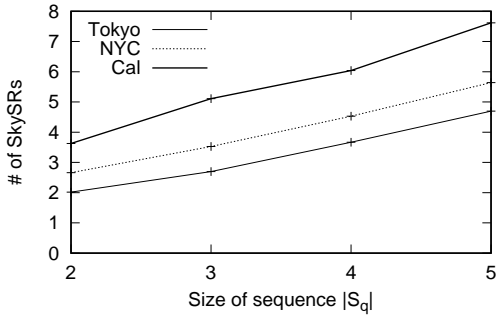}
	 	\vspace{-3mm}
	 	\caption{Number of SkySRs for various $|\mathbf{S}_q|$}
	 	\label{fig:SKY}
	 	\vspace{-5mm}   
\end{figure}

\subsection{Number of skyline sequenced routes}
Figure \ref{fig:SKY} shows the number of SkySRs obtained with each dataset for various $|\mathbf{S}_q|$.
As shown, the Cal dataset returns the largest number of SkySRs.
The response time and RSS obtained with the Tokyo and NYC datasets are much greater than the those of the Cal dataset, which implies that the number of SkySRs does not affect response time and RSS significantly. 
Moreover, if we use a complete real-world dataset, we may not require a ranking function because the number of SkySRs would be small.  

\subsection{Usecase}
\begin{figure}[t]
	 	\centering
	 	\includegraphics[width=0.9\linewidth]{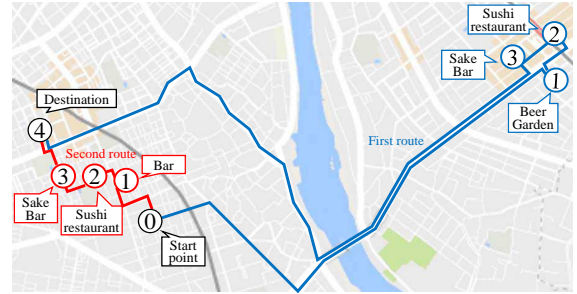}
	 	\caption{Visualization of routes in Tokyo: black circles (with 0 and 4) denote a start point and a destination, respectively. Blue and red circles denote sequences of PoIs for the first and second routes in Table \ref{table:result_tokyo}, respectively, and their numbers indicate the order of PoIs to be visited.}
	 	\label{fig:tokyo_route}
	 	 \vspace{-4mm}
\end{figure}

We show an example of SkySRs in Tokyo.
We assume that we plan to go to places for dinner and drinks. 
We want to visit a ``Beer garden'', a ``Sushi restaurant'',  and a ``Sake bar'' from our current location and finally go to our hotel.
Table \ref{table:result_tokyo} and Figure \ref{fig:tokyo_route} show two representative SkySRs from the four identified SkySRs. Note that the other two routes are similar to either of the representative routes.
In the Foursquare category trees, ``Bar'' includes ``Beer Garden'' and ``Sake bar'', and ``Japanese restaurant'' includes ``Sushi restaurant''.
Thus, we find routes using ``Bar'' and/or ``Japanese restaurant''.
The second route is much shorter than the first route that perfectly matches the user requirement, and the difference between them is only whether they pass a ``Bar'' or ``Beer garden''.
The best route depends on the users and situations (e.g., weather); thus, we confirm that SkySRs are useful to help users make decisions.

\begin{table}[t]
	\begin{center}
	\caption{Example SkySRs in Tokyo}
		\label{table:result_tokyo}
		\scriptsize	
		\begin{tabular}{|l|l|} \hline
Distance & Sequenced route	\\ \hline \hline
7451 meters & Beer Garden $\rightarrow$ Sushi Restaurant $\rightarrow$ Sake Bar \\
1295 meters& Bar $\rightarrow$ Sushi Restaurant $\rightarrow$ Sake Bar\\ \hline
		\end{tabular}
        \end{center}
\end{table}

\section{User Study}
\label{sec:usertest}
\begin{figure}[t]
 	 	\centering
	 	\includegraphics[width=0.9\linewidth]{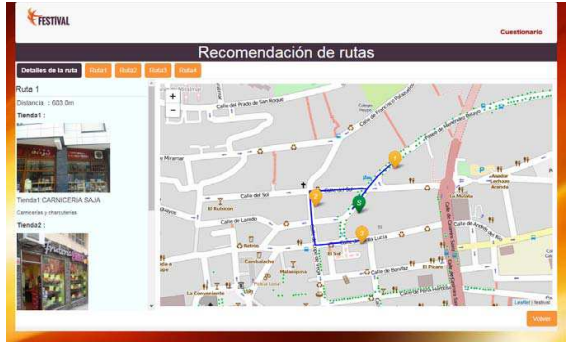}
	 	\caption{Screenshot of the prototype system}
	 	\label{fig:prototype}
	 	\vspace{-5mm}   
\end{figure}

We developed a prototype SkySR query service\footnote{https://ss.festival.ckp.jp/OuRouteSuggestion/dispSearchRoute/index. The default language is Spanish.} using OpenStreetMap and the Santander Open Data platform from Santander, Spain\footnote{http://datos.santander.es}. Figure \ref{fig:prototype} shows a screenshot of the prototype system, which outputs one of the SkySR route.
We performed a test in July, 2017. To gather users for this test, the Santander municipality arranged meetings with different groups of people to present the service: municipal staff (computing, convention and tourism municipal services), students from vocational training departments who are developing webpages and apps, and citizens.
We also provided a leaflet that shows the concept of the SkySR query and how to use the service.
In this test, users freely used the service and answered a questionnaire (25 respondents).
The questionnaire included the following three questions. 
\begin{itemize}
\item[{\bf Q1}] What do you think about this service? \\{\bf Answer.} 1. I love it, 2. I like it, 3. I do not like it. 
\item[{\bf Q2}] Would you recommend it to anyone? \\{\bf Answer.} 1. Yes, 2. Maybe, 3. No. 
\item[{\bf Q3}] Do you think that it is a good idea for the city: citizens, tourists, commercial sectors? \\{\bf Answer.} 1. Yes, 2. Maybe, 3. No.
\end{itemize}

\begin{figure}[t]
 	 	\centering
	 	\includegraphics[width=0.9\linewidth]{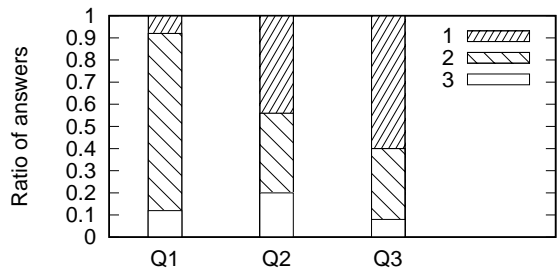}
	 	\vspace{-7mm}
	 	\caption{Ratios of answers for each question}
	 	\label{fig:usertest}
	 	\vspace{-3mm}   
\end{figure}

We summarize the ratios of answers for each question in Figure \ref{fig:usertest}. As shown, more than 80$\%$ of the users liked the service.
In addition, the questionnaire shows that the service is valuable for the city.
From the user experiment, we confirm that the SkySR query is useful for users and cities.


\section{Conclusion}
\label{sec:con}

In this paper, we have first introduced a semantic hierarchy for trip planning. We then proposed the skyline sequenced route (SkySR) query, which finds all preferred routes from a start point according to a user's PoI requirements.
In addition, we have proposed an efficient algorithm for the SkySR query, i.e., {\sf BSSR}, which simultaneously searches for all SkySRs by a single traversal of a given graph.
To optimize the performance of {\sf BSSR}, we proposed four optimization techniques. 
We evaluated the proposed approach using real-world datasets and demonstrated that it comprehensively outperforms naive approaches in terms of response time without increasing memory usage or sacrificing the exactness of the result. Moreover, we developed a SkySR query service using open data, and conducted a user test, which confirmed that SkySR queries are useful for both users and cities.

In future work, we would like to extend the proposed approach in several directions. First, because we assume a forest structure for the category classification in this paper, a more complex classification may provide better granularity. Second, because we have not used any preprocessing techniques such as indexing, we plan to propose a suitable preprocessing method for the SkySR query. Finally, although the SkySR query proposed in this paper considers two scores (length and category similarity), it could be extended to consider many attributes of a PoI (e.g., text, keywords, and ratings) and the cost/quality of a graph (e.g., route popularity, tolls, and the number of traffic lights).

\section*{Acknowledgement}
This research is partially supported by the Grant-in-Aid for
Scientific Research (A)(JP16H01722) and Grant-in-Aid for Young Scientists (B)(JP15K21069).

\bibliographystyle{ACM-Reference-Format}
\bibliography{camera-EDBT2018} 
\end{document}